\documentclass[aps,showpacs,floatfix,superscriptaddress,a4paper]{revtex4}
\usepackage{graphicx,amsmath,bbm,mathrsfs,amssymb,psfrag}

\newtheorem{theorem}{Theorem}
\newtheorem{corollary}{Corollary}
\newtheorem{lemma}{Lemma}
\newtheorem{definition}{Definition}
\newtheorem{proposition}{Proposition}
\newtheorem{example}{Example}
\newtheorem{remark}{Remark}

\newcommand{\qed}{\hfill $\Box$\medskip}
\newcommand{\proof}{\noindent{\bf Proof:\quad }}
\newcommand{\Tr}{{{\rm Tr}}}
\newcommand{\QI}{{{\rm Q}}}

\newcommand{\CI}{{\mathbb{C}}}

\newcommand{\id}{{{\rm id}}}
\newcommand{\rT}{{\rm T}}
\newcommand{\1}{{\mathbbm{1}}}

\newcommand{\cV}{{\mathcal{V}}}
\newcommand{\cQ}{{\mathcal{Q}}}
\newcommand{\cI}{{\mathcal{I}_{spec}}}
\newcommand{\hD}{\widehat{\Delta}}

\begin{document}

\title{Entanglement witnesses for a class of bipartite states of $n\times n$ qubits}

\author{Fabio Benatti}
\affiliation{Dipartimento di Fisica, Universit\`a degli Studi di
Trieste, I-34151 Trieste, Italy}
\affiliation{Istituto Nazionale di Fisica Nucleare, Sezione di
  Trieste, I-34151 Trieste, Italy}

\author{Mahya Karbalaii}
\affiliation{SISSA - Via Bonomea 265 - 34136, Trieste - Italy}

\begin{abstract}
We characterize  the positive maps  detecting the entangled bipartite states of $n\times n$ qubits that are diagonal with respect to the orthonormal basis constructed by tensor products of Pauli matrices acting on the totally symmetric state.
We then discuss the case $n=2$ for a class of states completely determined by the geometric patterns of subsets of a $16$ point lattice.
\end{abstract}

\maketitle

\section{Introduction}

Entangled quantum states are among the most important physical resources in the manifold applications of quantum information theory~\cite{Hororev}; from a  mathematical point of view, entanglement is closely related to positive and completely positive linear maps on operator algebras.
In the following, we shall consider finite dimensional bipartite quantum systems described by Hilbert spaces $\CI^{2^n}\otimes\CI^{2^n}$:  in this case, identifying entangled states is equivalent to sorting out positive maps $\Lambda: M_{2^n}(\CI)\mapsto M_{2^n}(\CI)$, which are not completely positive, such that these states do not remain positive under the action of the map $\id\otimes\Lambda$.
When $n=1$, all entangled states are detected by acting on just one of the two parties with the transposition~\cite{Horo,Per}. In low dimension, all states which remain positive under partial transposition (PPT) are automatically separable for positive maps always involve the transposition~\cite{Sto,Wor}. In higher dimension, this is not the case and there can be PPT entangled states~\cite{PHor}. Since a general characterization of positive maps is not available, a better control on both the entanglement of states and the positivity of maps  can only come by devising new positive maps that may detect the entanglement of at least particular classes of bipartite states~\cite{Kye,breuer,Hall,ChrKos1,ChrKos2,benatti0,benatti1,benatti2,ChrKos3}.
In this spirit, we consider in the following bipartite states that are diagonal with respect to the orthonormal basis generated by the action of tensor products of the form
$\1_{2^n}\otimes\sigma_{\vec{\mu}}$, $\sigma_{\vec{\mu}}=\otimes_{i=1}^n\sigma_{\mu_i}$, on the totally symmetric state $\vert\Psi^{2^n}_+\rangle\in  \CI^{2^n}\otimes\CI^{2^n}$.
We first characterize the structure of positive maps  detecting the entangled ones among them; then, we illustrate the result by examining some entanglement witnesses, already present in the literature~\cite{benatti0,benatti1,benatti2} for the case $n=2$, that is when the states correspond to normalized projections onto subspaces generated by orthogonal vectors of the form $\1_4\otimes\sigma_{\mu_1\mu_2}\vert\Psi^4_+\rangle\in\CI^{16}$. Finally, we show how, for this class of states being separable, entangled or PPT entangled are properties related to the geometric patterns of the subsets of $16$ square lattice points which identify them.

\section{Positive maps and entangled states}

In this section we resume some definitions and known facts about positive and completely positive maps on one hand and separable and entangled states, on the other hand (see for instance~\cite{karol,benattibook}).

Given the Hilbert space $\CI^{d_1}\otimes\CI^{d_2}$ of a bipartite systems consisting of two finite level systems, the states (density matrices) over it
separate into two sets.

\begin{definition}
\label{defsep}
Those density matrices on $\CI^{d_1}\otimes\CI^{d_2}$ that can be written as convex combinations of tensor products of states for the individual parties
\begin{equation}
\label{sep}
\rho=\sum_{i,j}\lambda_{ij}\,\rho^{(1)}_i\otimes\rho_j^{(2)}\ ,\qquad\lambda_{ij}\geq 0\ ,\quad\sum_{ij}\lambda_{ij}=1\ ,
\end{equation}
are called separable states and form a closed convex set.

The complementary set of those density matrices that cannot be written in the form of above is the set of entangled states.
\end{definition}

Let $M_d(\CI)$ be the algebra of $d\times d$ matrices acting on $\CI^d$.

\begin{definition}
\label{def1}
A linear map $\Lambda:M_{d_1}(\CI)\mapsto M_{d_2}(\CI)$ is said to be positive (P)  if it sends positive matrices into positive matrices.
Let $\id_n:M_n(\CI)\ni X\mapsto X$ denote the identity map on $M_n(\CI)$. Then, $\Lambda$ is said to be completely positive (CP) if $\id_n\otimes\Lambda$ is positive on $M_n(\CI)\otimes M_{d_1}(\CI)$ for all $n\geq 1$.
\end{definition}

Contrary to positive maps, the structure of completely positive maps is wholly determined by the following theorem.

\begin{theorem}
\label{KS}
A linear map $\Lambda:M_d(\CI)\mapsto M_d(\CI)$ is completely positive if and only if it decomposes as
$$
M_{d_1}(\CI)\ni X\mapsto\Lambda[X]=\sum_jL_j^\dag\,X\,L_j\in M_{d_2}(\CI)\ ,
$$
where $L_j:\CI^{d_2}\mapsto\CI^{d_1}$ are matrices such that the sum $\sum_jL_j^\dag L_j$ converges.
\end{theorem}

We fix an orthonormal basis $\{\vert i\rangle\}^{d}_{i=1}$ of $\CI^d$ and introduce the symmetric state
\begin{equation}
\label{symmstate1}
\vert \Psi^{d}_{+}\rangle=\frac{1}{\sqrt{d}} \sum^{d}_{i=1} \vert
i\rangle\otimes \vert i\rangle\in \CI^d\otimes \CI^d
\end{equation}
and the corresponding projector
\begin{equation}
\label{symmstate2}
P^{d}_{+}=\vert\Psi^{d}_{+}\rangle\langle\Psi^{d}_{+}\vert\in M_d(\CI)\otimes M_d(\CI)=M_{d^2}(\CI)\ .
\end{equation}

\begin{definition}
\label{DefChoi}
Given a positive map $\Lambda: M_{d_1}(\CI)\mapsto M_{d_2}(\CI)$, the matrix
\begin{equation}
\label{Choimatrix}
M_\Lambda:=\id_{d_1}\otimes\Lambda[P^{d_1}_+]\in M_{d_1}(\CI)\otimes M_{d_2}(\CI)=M_{d_1d_2}(\CI)
\end{equation}
is called the Choi matrix of $\Lambda$.
\end{definition}
\medskip

The positivity and complete positivity of a map $\Lambda$ can be deduced by inspecting its Choi matrix~\cite{Choi,Jam}.

\begin{proposition}
\label{prop1}
A map $\Lambda: M_{d_1}(\CI)\mapsto M_{d_2}(\CI)$ is positive if and only if its Choi matrix is block-positive, namely if
\begin{equation}
\label{blockpos}
\langle\varphi\vert\,\Lambda[\vert\psi\rangle\langle\psi\vert]\,\vert\varphi\rangle\geq0\Longleftrightarrow
\langle\psi^*\otimes\varphi\vert\, M_\Lambda\,\vert\psi^*\otimes\varphi\rangle\geq 0
\end{equation}
for all $\vert\varphi\rangle\in\CI^{d_2}$ and $\vert\psi\rangle\in \CI^{d_1}$, where $\vert\psi^*\rangle$ denotes the conjugate vector with respect to a fixed orthonormal basis.

A map $\Lambda: M_{d_1}(\CI)\mapsto M_{d_2}(\CI)$ is completely positive map if and only if its Choi matrix is positive.
\end{proposition}

The transposition map $T:M_d(\CI)\mapsto M_d(\CI)$ such that, in a chosen representation,
$$
M_d(\CI)\ni X\mapsto T[X]\in M_d(\CI)\ ,\quad (T[X])_{ij}=X_{ji}
$$
is the prototype of a positive, but not completely positive map. Indeed, its Choi matrix
$$
{\rm id_d}\otimes T[P^d_+]=\frac{1}{d}\sum_{i,j=1}^d\vert i\rangle\langle j\vert\otimes \vert j\rangle\langle  i\vert
$$
is proportional to the flip operator $V\vert\psi\otimes\phi\vert=\vert\phi\otimes\psi\rangle$ that has eigenvalues $\pm1$.

Since the subset of bipartite separable states  is convex and closed, the Hahn-Banach theorem can be used to separate it from  any entangled state $\rho_{ent}$  by an hyperplane; this latter is in turn represented by a matrix $M\in M_d(\CI)\otimes M_d(\CI)$ such that
${\rm Tr}(M\, \rho_{ent})\,<\,0\,\leq\,{\rm Tr}(\rho\,M)$ for all separable states $\rho$.
The matrix $M$ must thus be block-positive, thence identified by a positive map $\Lambda$ such that~\cite{Horo}
\begin{equation}
\label{HB}
{\rm Tr}\left({\rm id}_d\otimes\Lambda[P^d_+]\, \rho_{ent}\right)\,<\,0\,\leq\,{\rm Tr}\left(\rho\,{\rm id}_d\otimes\Lambda[P^d_+]\right)\qquad\forall \rho\in\mathcal{S}_{sep}\ .
\end{equation}

The possibility of witnessing the entanglement of a bipartite state $\rho\in M_d(\CI)\otimes
M_d(\CI)$ by means of a positive map  is the content of~\cite{Horo,Per}
\medskip

\begin{proposition}
\label{prop2}
A state $\rho\in M_{d_1d_2}(\CI)$ is separable if and only if
$$
(\id_{d_1}\otimes\Lambda)[\rho]\geq0\ ,
$$
for all positive maps $\Lambda: M_{d_2}(\CI)\mapsto M_{d_1}(\CI)$.
Equivalently,  $\rho\in M_{d_1d_2}(\CI)$ is entangled if and only if there exists a positive map $\Lambda: M_{d_1}(\CI)\mapsto M_{d_2}(\CI)$ with $M_\Lambda={\rm id}_{d_1}\otimes\Lambda[P^{d_1}_+]\in M_{d_1d_2}(\CI)$ such that
$\Tr\Big(M_\Lambda\,\rho\Big)<0$.
\end{proposition}

In the following we shall freely call \textit{entanglement witnesses} both positive maps $\Lambda$ such that ${\rm id}_d\otimes\Lambda[\rho_{ent}]\ngeq0$ and their Choi matrices $M_\Lambda$.
\medskip

\begin{remark}
\label{rem2}
In general, each entangled state $\rho\in M_{d_1d_2}(\CI)$ has its own entanglement witnesses. However, when $d_1d_2\leq6$, namely in the case of two qubits or one qubit and one qutrit, all positive maps $\Lambda: M_{d_1}(\CI)\mapsto M_{d_2}(\CI)$, $d_1d_2\leq6$, are decomposable in the form $\Lambda=\Lambda_1+\Lambda_2\circ T$, where $\Lambda_{1,2}$ are completely positive maps from  $M_{d_1}(\CI)$ to  $M_{d_2}(\CI)$ and $T$ is the transposition map on $M_{d_1}(\CI)$~\cite{Sto,Wor} .
It thus follows that, in such lower dimensional cases, the only ones where positive maps have a definite structure, a bipartite state is entangled if and only if it does not remain positive under partial transposition.
In higher dimension, that is for $d_1d_2\geq8$, not all positive maps are decomposable; consequently there can be bipartite entangled states that remain positive under partial transposition (PPT entangled states) \cite{PHor,Hororev}. Furthermore, the lack of a complete characterization of generic positive maps entails a lack of control on the separability of generic bipartite states. Therefore, a deeper understanding can only be gathered by seeking entanglement witnesses adapted to certain specific classes of states in the hope that the accumulated phenomenology might help to shed light both on the phenomenon of entanglement and on the characterization of positive maps.
\end{remark}

One useful technical step that will be used in the following is~\cite{Stoermer}

\begin{proposition}
\label{Stormer}
Any positive map $\Lambda: M_{d_1}(\CI)\mapsto M_{d_2}(\CI)$ can be written as
\begin{equation}
\label{Stormer1}
\Lambda=\mu\left ({\rm Tr}\,-\,\Lambda^{CP}\right)\ ,
\end{equation}
where $\mu$ is  a positive proportionality factor, $\rm Tr$ is the trace operation such that
\begin{equation}
\label{Trop}
X\in M_{d_1}(\CI)\mapsto {\rm Tr}[X]={\rm Tr}(X)\,\1_{d_2}\ ,
\end{equation}
while
$\Lambda^{CP}: M_{d_1}(\CI)\mapsto M_{d_2}(\CI)$ is a suitable CP map.
\end{proposition}

\begin{example}
\label{ex1}
As a concrete example that will turn useful in the following, we consider $d_1=d_2=4$, that is two parties each consisting of two qubits; let $\sigma_{\alpha}$, $\alpha=0,1,2,3$ denote the Pauli matrices with $\sigma_{0}=\1_2$, the $2\times 2$ identity matrix.
We choose the representation where $\sigma_3$ is diagonal; then, $T[\sigma_\alpha]=\sigma_\alpha$ if $\alpha\neq 2$, otherwise $T[\sigma_2]=-\sigma_2$. Therefore, using the
anti-commutation relations $\sigma_i\sigma_j=-\sigma_j\sigma_i$, $i\neq j$, the transposition acts as
\begin{equation}
\label{transp2}
M_2(\CI)\ni X\mapsto T[X]=\frac{1}{2}\sum_{\alpha=0}^3\,\varepsilon_\alpha\,S_{\alpha}[X]\ ,\quad S_{\alpha}[X]=\sigma_\alpha\,X\,\sigma_\alpha\ ,\quad \varepsilon_\alpha=(1,1,-1,1)\ .
\end{equation}
The extension to $M_4(\CI)$ is straightforward:
\begin{equation}
\label{transp4}
M_4(\CI)\ni X\mapsto T[X]=\frac{1}{4}\sum_{\alpha,\beta=0}^3\,\varepsilon_\alpha\varepsilon_\beta\,S_{\alpha\beta}[X]\ ,\quad
S_{\alpha\beta}[X]=\sigma_{\alpha\beta}\,X\,\sigma_{\alpha\beta}\ ,\quad \sigma_{\alpha\beta}=\sigma_{\alpha}\otimes\sigma_\beta\ .
\end{equation}
Notice that the transposition is not written in the Kraus-Stinespring form~(\ref{KS}) since $\varepsilon_2=-1$ in (\ref{transp2}), while in (\ref{transp4}) the products $\varepsilon_\alpha\varepsilon_\beta=-1$ whenever $\alpha\neq\beta=2$ or $\beta\neq\alpha=2$.

On the other hand,  the CP trace map~(\ref{Trop}) on $M_2(\CI)$ can be put in the form~(\ref{KS})
\begin{equation}
\label{Tr2}
{\rm Tr}[X]=\frac{1}{2}\sum_{\alpha=0}^3S_\alpha[X]\ ,\qquad X\in M_2(\CI)\ ,
\end{equation}
which gives ${\rm Tr}[\sigma_\alpha]=\delta_{\alpha 0}\1_2$. As the trace map on $M_2(\CI)$, also its extension to $M_4(\CI)$  can be written in the Kraus-Stinespring form~(\ref{KS}):
\begin{equation}
\label{Tr4}
M_4(\CI)\ni X\mapsto{\rm Tr}[X]=\frac{1}{4}\sum_{\alpha,\beta=0}^3\,S_{\alpha\beta}[X]\ .
\end{equation}

 When $\Lambda$ is the transposition map on $M_4(\CI)$, the CP maps $\Lambda^{CP}$ in~(\ref{Stormer1}) are easily found: the linear map
$$
{\rm Tr}-\frac{1}{\mu}\,T=\frac{1}{4}\sum_{\mu,\nu=0}^3\,\left(1-\frac{\varepsilon_\alpha\varepsilon_\beta}{\mu}\right)\, S_{\alpha\beta}\ ,
$$
is of the form~(\ref{KS}), thus CP, if and only if $\mu\geq 1$. The smallest choice, $\mu=1$, yields the completely positive map
\begin{equation}
\label{CPTRansp}
\Lambda^{CP}=\frac{1}{2}\Big(\sum_{\alpha\neq 2}\,S_{\alpha2}\,+\,\sum_{\beta\neq 2}\,S_{2\beta}\Big)\ .
\end{equation}
\end{example}

\subsection{$\sigma$-diagonal and lattice states}

We shall consider a bipartite system consisting of two parties in turn comprising $n$ qubits; the corresponding matrix algebra $M_{2^{2n}}(\CI)$ is linearly spanned by $4^n$ tensor products of the form  $\sigma_{\vec{\mu}}:=\otimes_{i=1}^n\sigma_{\mu_i}=\sigma_{\mu_1}\otimes\sigma_{\mu_2}\otimes\cdots\sigma_{\mu_n}$.

Given the totally symmetric vector~(\ref{symmstate1}) with $d=2^n$, $\vert \Psi^{2^n}_+\rangle\in\CI^{4^n}$, the vectors
\begin{equation}
\label{ONV}
\vert\Psi_{\vec{\mu}}\rangle:=\1_{2^n}\otimes\sigma_{\vec{\mu}}\vert\Psi^{2^n}_+\rangle\in \CI^{2^n}\otimes\CI^{2^n} \ ,
\end{equation}
form orthogonal projectors
\begin{equation}
\label{ONB16}
P_{\vec{\mu}}:=\vert\Psi_{\vec{\mu}}\rangle\langle\Psi_{\vec{\mu}}\vert
=(\1_{2^n}\otimes\sigma_{\vec{\mu}})\vert\Psi^{2^n}_{+}\rangle\langle\Psi^{2^n}_{+}\vert(\1_{2^n}\otimes\sigma_{\vec{\mu}})\ .
\end{equation}
Orthonormality follows since
$$
\langle\Psi_{\vec{\nu}}\vert\Psi_{\vec{\mu}}\rangle=\langle\Psi^{2^n}_+\vert\1_{2^n}\otimes\sigma_{\vec{\mu}}\sigma_{\vec{\mu}}\vert\Psi^{2^n}_+\rangle=
\frac{1}{2^n}{\rm Tr}\left(\sigma_{\vec{\nu}}\sigma_{\vec{\mu}}\right)
=\frac{1}{2^n}\prod_{i=1}^n{\rm Tr}\left(\sigma_{\nu_i}\sigma_{\mu_i}\right)=\prod_{i=1}^n\delta_{\nu_i\mu_i}\ .
$$

\begin{definition}
\label{s-diagstates}
The class of bipartite states we shall study will consist of states of the form
\begin{equation}
\label{diagstates}
\rho=\sum_{\vec{\mu}}\  r_{\vec{\mu}}\ P_{\vec{\mu}}\ ,\qquad 0\leq r_{\vec{\mu}}\leq 1\ ,\quad\sum_{\vec{\mu}}\ r_{\vec{\mu}}=1\ ,
\end{equation}
that is diagonal with respect to the chosen orthonormal basis $\{\vert\Psi_{\vec{\mu}}\rangle\}_{\vec{\mu}\in\{0,1,2,3\}^n}$ in $\CI^{2^n}\otimes\CI^{2^n}$: we shall call them $\sigma$-diagonal states.
\end{definition}

A particular sub-class of states of two pairs of two qubits, $n=2$ and $\sigma_{\vec{\mu}}=\sigma_\alpha\otimes\sigma_\beta=\sigma_{\alpha\beta}\in M_4(\CI)$, were considered in~\cite{benatti0,benatti1,benatti2} and called lattice states. When they are PPT, the entanglement properties of such states
cannot be ascertained by standard methods, like for instance the reshuffling criterion~\cite{karol,mahya}, and new methods had to be devised.
We shall tackle these states again in the following.
\medskip

\begin{definition}
\label{def:lattstate}
Taking a subset $I\subseteq L_{16}$ of cardinality $N_{I}$, then the corresponding lattice state $\rho_{I}$ is defined by:
\begin{equation}
\label{lattstate}
\rho_{I}=\frac{1}{N_{I}}\sum_{\alpha,\beta\in I }P_{\alpha\beta}\ .
\end{equation}
Let $L_{16}$ denote the set of pairs $(\alpha,\beta)$, where $\alpha$ and $\beta$ run from $0$ to $3$: it corresponds to a $4\times 4$ square lattice, whereas the subsets
\begin{equation}
\label{ColRows}
C_{\alpha}:=\{(\alpha,\beta)\in L_{16} : \beta=0,1,2,3\}\quad and \quad R_{\beta}:=\{(\alpha,\beta)\in L_{16} : \alpha=0,1,2,3\}
\end{equation}
correspond to the columns and rows of the lattice, respectively.
\end{definition}

If compared with those in (\ref{diagstates}), the lattice  states are uniformly distributed over chosen subsets $I\subseteq L_{16}$, and thus completely characterized by them.
Let us consider the lattice states associated with the following subsets:
\begin{eqnarray}
\label{Ex1}
N_I=8:\qquad
\begin{array}{c|c|c|c|c}
               3 & \quad\! & \times & \times &\quad\!   \\
               \hline
               2 & \times & \quad\! &  \quad\! &\times  \\
               \hline
               1 & \quad\! & \times& \times  &\quad\!   \\
               \hline
               0 & \times & \quad\! & \times  &   \\
               \hline
                 & 0 & 1 & 2 & 3
\end{array}
\qquad N_I=5:\qquad
\begin{array}{c|c|c|c|c}
               3 & \quad\! & \quad\! & \times &\quad\!   \\
               \hline
               2 & \times & \quad\! &\quad\! &\times  \\
               \hline
               1 & \quad & \quad\! & \times  &\quad\!   \\
               \hline
               0 & \quad\! & \quad\! & \quad\!  &\times   \\
               \hline
                 & 0 & 1 & 2 & 3
\end{array}\qquad .
\end{eqnarray}

The following proposition shows that these states do not remain positive under partial transposition and are therefore entangled: indeed, positivity under partial transposition (PPT-ness) of lattice states $\rho_I$  is completely characterized by the geometry of $I$~\cite{benatti0}.

\begin{proposition}
\label{PPT}
A necessary and sufficient condition for a lattice state $\rho_I$ to be PPT is that for every
$(\alpha,\beta)\in L_{16}$ the number of points on $C_\alpha$ and $R_\beta$  belonging to
$I$ and different from $(\alpha, \beta)$ be not greater than $N_I/2$. In terms of the characteristic functions $\chi_I(\alpha,\beta) = 1$ if
$(\alpha, \beta)\in I$, $= 0$ otherwise, a lattice state $\rho_I$ is PPT if and only if for all $(\alpha, \beta)\in L_{16}$:
$$
\sum_{0=\delta\neq\beta}^3\chi_I(\alpha, \delta) + \sum_{0=\delta\neq\alpha}^3\chi_I({\delta,\beta})\leq\frac{N_I}{2}\ .
$$
\end{proposition}
\medskip

In the first pattern of~(\ref{Ex1}), the row and column passing through the point $(2,2)\notin I$ contains $5>8/4=2$ points in $I$, in the second one $4>5/2$ points.

By the same criterion, the following two states are instead PPT,

\begin{eqnarray}
\label{Ex2}
N_I=6:\qquad
\begin{array}{c|c|c|c|c}
               3 & \quad\! & \quad\! & \times &\times   \\
               \hline
               2 & \times & \quad\! &  \quad\! &\times  \\
               \hline
               1 & \quad\! & \times & \quad\!  &\times   \\
               \hline
               0 & \quad\! & \quad\! & \quad\!  &   \\
               \hline
                 & 0 & 1 & 2 & 3
\end{array}
\qquad N_I=8:\qquad
\begin{array}{c|c|c|c|c}
               3 & \quad\! & \times & \times &\times   \\
               \hline
               2 & \times & \quad\! &\times &\times  \\
               \hline
               1 & \quad\! & \quad\! & \times  &\times   \\
               \hline
               0 & \quad\! & \quad\! & \quad\!  &\quad\!   \\
               \hline
                 & 0 & 1 & 2 & 3
\end{array}\qquad .
\end{eqnarray}
They need not be separable as in lower dimension; indeed, a sufficient criterion devised in~\cite{benatti1} and based on entanglement witnesses proposed in~\cite{breuer,Hall}, show them to be entangled.

\begin{proposition}
\label{PPT2}
A sufficient condition for a PPT lattice state $\rho_I$ to be entangled is that there exists at
least a pair $(\alpha, \beta)\in L_{16}$ not belonging to $I$ such that only one point on $C_{\alpha}$ and $R_\beta$ belongs to $I$.
Equivalently, $\rho_I$ is entangled if there exists a pair $(\alpha, \beta)\in L_{16}$, $  (\alpha, \beta) \notin I$, such that
$$
\sum_{0=\delta\neq\beta}^3\chi_I(\alpha, \delta) + \sum_{0=\delta\neq\alpha}\chi_I(\delta, \beta) = 1 \quad .
$$
\end{proposition}
\medskip

In both patterns of the states in Example~\ref{Ex2}, it is the point $(0,0)\notin I$ which satisfies the sufficient criterion.
Unfortunately, this criterion fails in the case of the PPT lattice state characterized by the following subset
\begin{eqnarray}
\label{Ex3}
N_I=10:\qquad
\begin{array}{c|c|c|c|c}
               3 & \times & \quad & \quad\! &\times  \\
               \hline
               2 & \quad\! & \times &\times &\quad\!  \\
               \hline
               1 & \times & \times & \quad\!  &\times   \\
               \hline
               0 & \times & \times & \quad\!  &\times   \\
               \hline
                 & 0 & 1 & 2 & 3
\end{array}\qquad .
\end{eqnarray}
Indeed, the only candidate point to fulfil the criterion in Proposition~\ref{PPT2} is $(2,2)$; however, it belongs to $I$.
Luckily, a refined criterion~\cite{benatti2} based on~\cite{breuer} shows it to be entangled.

\begin{proposition}
\label{PPT3}
A $PPT$ lattice state $\rho_I$
is entangled  if there exists $(\mu,\nu)\in L_{16}$ such that the quantity
$$
k_I^{\mu\nu}=
\sum_{\alpha\neq\nu\oplus2}\chi_I(\alpha, \nu \oplus 2) + \sum_{\beta\neq\mu\oplus2}\chi_I(\mu \oplus 2,\beta)\ ,
$$
where $\oplus$ denotes summation modulo $4$,
is such that $k_I^{\mu\nu}= 1$ for a column $C_{\mu\oplus 2}$ and a row $R_{\nu\oplus 2}$, independently of whether $(\mu\oplus 2, \nu\oplus 2)$ belongs to $I$ or not.
\end{proposition}
\medskip

Thus, the state in~(\ref{Ex3}) fulfils the sufficient condition $k_I^{00}=1$.

These examples suggest a relation between the structure of the subsets that define the lattice states and their entanglement properties;
in the following we will try to clarify this correspondence.

\section{Entanglement detection for $\sigma$-diagonal states}

In this section, we will show that witnesses of the entanglement of states of the form~(\ref{diagstates}) can be sought within a particular subclass of positive maps from $M_{2^n}(\CI)$ onto $M_{2^n}(\CI)$.

Any linear map on $\Lambda: M_d(\CI)\mapsto M_d(\CI)$ can be written as~\cite{benattibook}:
$$
M_d(\CI)\ni X\mapsto \Lambda[X]=\sum^{d^{2}-1}_{i,j=0}\lambda_{ij}\ G_i\,X\,G^{\dagger}_{j}\ ,
$$
where the matrices $G_{i}\in M_d(\CI)$ constitute a so-called Hilbert-Schmidt orthonormal basis in $M_d(\CI)$, namely ${\rm Tr}(G_i^\dag\,G_j)=\delta_{ij}$ and the coefficients $\lambda_{ij}$ are complex numbers.

In the present  case, the normalized  tensor products $\displaystyle\frac{\sigma_{\vec{\mu}}}{\sqrt{2^n}}$ constitute such a basis in $M_{2^n}(\CI)$, whence linear maps $\Lambda:M_{2^n}(\CI)\mapsto M_{2^n}(\CI)$ can be expressed as
\begin{equation}
\label{genexp}
M_{2^n}(\CI)\ni X\mapsto\Lambda[X]=\sum_{\vec{\mu},\vec{\nu}}\,\lambda_{\vec{\mu}\vec{\nu}}\ S_{\vec{\mu}\vec{\nu}}[X]\ ,\qquad
S_{\vec{\mu}\vec{\nu}}[X]=\sigma_{\vec{\mu}}\,X\,\sigma_{\vec{\nu}}\ .
\end{equation}

The next one is a simple observation based on~(\ref{ONV}) and~(\ref{ONB16}).

\begin{lemma}
\label{lemma1}
A $\sigma$-diagonal state $\rho=\sum_{\vec{\mu}}\ r_{\vec{\mu}}\ P_{\vec{\mu}}$ is entangled if and only if there exists a positive map $\Lambda$ as in ~(\ref{genexp}) such that
$\sum_{\vec{\mu}}\ \lambda_{\vec{\mu}\vec{\mu}}\, r_{\vec{\mu}}\,<\,0$.
\end{lemma}

\noindent
\proof
Because of Proposition~\ref{prop2} and of the orthogonality of the vectors $\vert\Psi_{\vec{\mu}}\rangle$ in~(\ref{ONV}), $\rho=\sum_{\vec{\mu}}\ r_{\vec{\mu}}\ P_{\vec{\mu}}$ is entangled if and only if
$$
{\rm Tr}\left(\id_{2^n}\otimes\Lambda[P^{2^n}_+]\,\rho\right)=\sum_{\vec{\mu},\vec{\nu}}\ \lambda_{\vec{\mu}\vec{\nu}}\ \langle\Psi_{\vec{\nu}}\vert\rho\vert\Psi_{\vec{\mu}}\rangle=
\sum_{\vec{\mu}}\ \lambda_{\vec{\mu}\vec{\mu}}\, r_{\vec{\mu}}\,<\, 0\ ,
$$
for some positive map $\Lambda: M_{2^n}(\CI)\mapsto M_{2^n}(\CI)$.
\qed
\medskip

The lemma indicates that the class of diagonal positive maps of the form
$\Lambda=\sum_{\vec{\mu}}\,\lambda_{\vec{\mu}}\ S_{\vec{\mu}\vec{\mu}}$
might suffice to witness the entanglement of states of the form $\rho=\sum_{\vec{\mu}}\ r_{\vec{\mu}}\ P_{\vec{\mu}}$.
What we need is the following result.

\begin{lemma}
\label{lemma2}
Given a positive map of the form $\Lambda=\sum_{\vec{\mu},\vec{\nu}}\,\lambda_{\vec{\mu}\vec{\nu}}\ S_{\vec{\mu}\vec{\nu}}$ , the diagonal map
$\Lambda_{diag}=\sum_{\vec{\mu}}\,\lambda_{\vec{\mu}\vec{\mu}}\ S_{\vec{\mu}\vec{\mu}}$ is also positive.
\end{lemma}

\proof
From Proposition~\ref{prop1}, the positivity of $\Lambda$ is equivalent to
\begin{equation}
\label{posconst1}
\sum_{\vec{\mu},\vec{\nu}}\lambda_{\vec{\mu}\vec{\nu}}\,\langle\varphi\vert
\sigma_{\vec{\mu}}\vert\psi\rangle\langle\psi\vert\sigma_{\vec{\nu}}\vert\varphi\rangle\,\geq\, 0\qquad \forall\  \vert\psi\rangle,\vert\varphi\rangle\in \CI^{2^n}\quad .
\end{equation}
Given a pair $\vert\psi\rangle,\vert\varphi\rangle\in \CI^{2^n}$, consider another pair
$\vert\psi_{\delta_i}\rangle=\sigma_{\delta_i}\vert\psi\rangle$, $\vert\varphi_{\delta_i}\rangle=\sigma_{\delta_i}\vert\varphi\rangle$,
where $\sigma_{\delta_i}$ denotes the tensor product $\sigma_{\vec{\alpha}}$ where $\alpha_j=0$ for $j\neq i$ and $\alpha_i=\delta_i\neq 0$.
Inserting the new pair into~(\ref{posconst1}), we get
\begin{equation}
\label{posconst2}
\sum_{\vec{\mu},\vec{\nu}}\ \lambda_{\vec{\mu}\vec{\nu}}\ \langle\varphi\vert\bigotimes_{j=1}^{i-1}\sigma_{\mu_j}\otimes
\,\Big(\sigma_{\delta_i}\sigma_{\mu_i}\sigma_{\delta_i}\Big)\,\bigotimes_{j=i+1}^{n}\sigma_{\mu_j}\vert\psi\rangle\langle\psi\vert
\bigotimes_{j=1}^{i-1}\sigma_{\nu_j}\otimes\,\Big(\sigma_{\delta_i}\sigma_{\nu_i}\sigma_{\delta_i}\Big)\,
\bigotimes_{j=i+1}^{n}\sigma_{\nu_j}\vert\varphi\rangle\,\geq\, 0\ .
\end{equation}
Consider $\mu_i\neq\nu_i$ and $\nu_i\neq 0$; because of the Pauli algebraic relations, one can always choose $\sigma_{\delta_i}$ in such a way that
$$
\sigma_{\delta_i}\sigma_{\mu_i}\sigma_{\delta_i}=\sigma_{\mu_i}\qquad\hbox{and}\qquad
\sigma_{\delta_i}\sigma_{\nu_i}\sigma_{\delta_i}=-\sigma_{\nu_i}\ ,
$$
whence all the terms in~(\ref{posconst2}) corresponding to the chosen pair of indices $(\mu_i,\nu_i)$ contribute with
$$
-\,\lambda_{\vec{\mu}\vec{\nu}}\,\langle\varphi\vert\bigotimes_{j=1}^{i-1}\sigma_{\mu_j}\otimes
\,\sigma_{\mu_i}\,\bigotimes_{j=i+1}^{n}\sigma_{\mu_j}\vert\psi\rangle\langle\psi\vert
\bigotimes_{j=1}^{i-1}\sigma_{\nu_j}\otimes\,\sigma_{\nu_i}\,
\bigotimes_{j=i+1}^{n}\sigma_{\nu_j}\vert\varphi\rangle\ .
$$
If $\nu_i=0$, the previous works with $\mu_i\neq 0$. Then, adding inequalities~(\ref{posconst1}) and~(\ref{posconst2}) yields
$$
\sum_{\vec{\mu},\vec{\nu}\,:\, \mu_i=\nu_i}\ \lambda_{\vec{\mu}\vec{\nu}}\ \langle\varphi\vert
\sigma_{\vec{\mu}}\vert\psi\rangle\langle\psi\vert\sigma_{\vec{\nu}}\vert\varphi\rangle\,\geq\,0\quad .
$$
By iterating the argument for all pairs $(\mu_i,\nu_i)$, one cancels all contributions from $\mu_i\neq \nu_i$ and remains with
\begin{equation}
\label{posconst3}
\sum_{\vec{\mu}}\ \lambda_{\vec{\mu}\vec{\mu}}\ \langle\varphi\vert
\sigma_{\vec{\mu}}\vert\psi\rangle\langle\psi\vert\sigma_{\vec{\mu}}\vert\varphi\rangle\, \geq\, 0\qquad\forall\ \vert\psi\rangle,\vert\varphi\rangle\in \CI^{2^n}\quad .
\end{equation}
This, by Proposition~\ref{prop1} amounts to the positivity of the diagonalized map $\Lambda_{diag}=\sum_{\vec{\mu}}\lambda_{\vec{\mu}\vec{\mu}}\,S_{\vec{\mu}\vec{\mu}}$.
\qed
\medskip

The previous result allows us to conclude with

\begin{proposition}
\label{diagprop}
Entangled $\rho=\sum_{\vec{\mu}}\ r_{\vec{\mu}}\, P_{\vec{\mu}}$ can be witnessed by diagonal positive maps
$\Lambda=\sum_{\vec{\mu}}\,\lambda_{\vec{\mu}\vec{\mu}}\ S_{\vec{\mu}\vec{\mu}}$.
\end{proposition}
\medskip

\proof
By the previous lemma, diagonalizing a positive map $\Lambda=\sum_{\vec{\mu},\vec{\nu}}\,\lambda_{\vec{\mu}\vec{\nu}}\ S_{\vec{\mu}\vec{\nu}}$ always yields
a positive map $\Lambda_{diag}=\sum_{\vec{\mu}}\,\lambda_{\vec{\mu}\vec{\mu}}\ S_{\vec{\mu}\vec{\mu}}$. Then, from Lemma~\ref{lemma1} it follows that either the entanglement of $\rho$ is witnessed by an already diagonal map or, if by a non-diagonal one, also  by
the map obtained by diagonalizing the latter.
\qed
\medskip

\begin{remark}
\label{rem3}
The above Proposition states that entangled states of the form~(\ref{diagstates}) can be witnessed by diagonal maps $\Lambda_{diag}=\sum_{\vec{\mu}}\,\lambda_{\vec{\mu}}\ S_{\vec{\mu}\vec{\mu}}$; the main problem is of course how to characterize the coefficients $\lambda_{\vec{\mu}}$ which cannot be all positive in such a way that~(\ref{posconst3}) be satisfied and thus $\Lambda_{diag}$ be positive.
The following lemma transforms that condition into a condition on the positive coefficients of a completely positive map associated to $\Lambda$ by means of
Proposition \ref{Stormer}.
\end{remark}

\begin{lemma}
\label{lemmaStormer}
All diagonal maps $\Lambda:M_{2^n}(\CI)\mapsto M_{2^n}(\CI)$, are positive if and only if
\begin{equation}
\label{Stormer3}
\Lambda_{diag}=\mu\sum_{\vec{\mu}}\Big(\frac{1}{2^n}\,-\,\lambda_{\vec{\mu}}\Big)\,S_{\vec{\mu}\vec{\mu}}\ ,\quad\mu>0\ ,
\end{equation}
where the coefficients $\lambda_{\vec{\mu}}$ must be positive and satisfy
\begin{equation}
\label{crit1}
\sum_{\vec{\mu}}\ \lambda_{\vec{\mu}}\,\left|\langle\varphi\vert\sigma_{\vec{\mu}}\vert\psi\rangle\right|^2\,\leq1\, \qquad\forall\  \vert\psi\rangle,\vert\varphi\rangle\in\CI^{2^n}\ .
\end{equation}
\end{lemma}

\proof
Using Proposition~\ref{Stormer}, all diagonal positive maps $\Lambda_{diag}$ can be related to completely positive diagonal maps
$$
\Lambda_{diag}^{CP}=\sum_{\vec{\mu}}\ \lambda_{\vec{\mu}}\,S_{\vec{\mu}\vec{\mu}}\ ,\qquad \lambda_{\vec{\mu}}\geq 0\ ,
$$
via the relation (\ref{Stormer1}) which,
extending the expression~(\ref{Tr4}) for the trace operation to $M_{2^n}(\CI)$ as follows
$$
{\rm Tr}=\frac{1}{2^n}\sum_{\vec{\mu}}\ S_{\vec{\mu}\vec{\mu}}\ ,
$$
can be recast in the form (\ref{Stormer3}). Then, asking for positivity implies, according to~(\ref{blockpos}),
\begin{equation}
\nonumber \label{Stormer4}
\langle\varphi\vert\Big({\rm Tr}-\Lambda^{CP}_{diag}\Big)[\vert\psi\rangle\langle\psi\vert]\vert\varphi\rangle=
1\,-\,\sum_{\vec{\mu}}\ \lambda_{\vec{\mu}}\,\left|\langle\varphi\vert\sigma_{\vec{\mu}}\vert\psi\rangle\right|^2\,\geq \,0\qquad\forall\ \vert\varphi\rangle,\vert\psi\rangle\in\CI^{2^n}\quad .
\end{equation}
\qed
\medskip

Using the previous lemma, we get necessary and sufficient conditions for the separability of $\sigma$-diagonal states.

\begin{proposition}
\label{propdiagst}
A  $\sigma$-diagonal  state $\rho=\sum_{\vec{\mu}}\ r_{\vec{\mu}}\,P_{\vec{\mu}}$ is separable if and only if for all sets of $4^n$ positive real numbers $\lambda_{\vec{\mu}}\geq 0$,
satisfying criterion (\ref{crit1}), it holds that
\begin{equation}
\sum_{\vec{\mu}}\ \lambda_{\vec{\mu}}\,r_{\vec{\mu}}\,\leq\,\frac{1}{2^n}\quad  .
\label{crit2}
\end{equation}
Otherwise, if a set of $4^n$ positive real numbers $\lambda_{\vec{\mu}}\geq 0$ can be found that satisfy~(\ref{crit1}) and such that
\begin{equation}
\sum_{\vec{\mu}}\ \lambda_{\vec{\mu}}\,r_{\vec{\mu}}\,>\,\frac{1}{2^n}\quad  ,
\label{crit3}
\end{equation}
then the $\sigma$-diagonal state $\rho=\sum_{\vec{\mu}}\ r_{\vec{\mu}}\,P_{\vec{\mu}}$ is entangled.
\end{proposition}

\proof
The result follows by inserting the expression  of the coefficients of diagonal positive maps resulting from (\ref{Stormer3})  into the condition
$\sum_{\vec{\mu}}\lambda_{\vec{\mu}\vec{\mu}}\,r_{\vec{\mu}}<0$ of Lemma \ref{lemma1}.
\qed
\medskip

Before tackling the case of lattice states, that is of $\sigma$-diagonal states with $n=2$, as a simple application, we consider the case of only two qubits, $n=1$, for which we know PPT-ness to coincide with separability.

\begin{example}
\label{ex2}
In the case $n=2$, $\sigma$-diagonal states have the form $\rho=\sum_{\mu=0}^3r_\mu P_\mu$, where $r_\mu\geq0$, $\sum_{\mu=0}^3r_\mu=1$ and
the $P_{\mu}$'s project onto the Bell states
\begin{eqnarray}
\nonumber
&&
\vert\Psi_0\rangle=\vert\Psi_+^2\rangle=\frac{\vert00\rangle+\vert11\rangle}{\sqrt{2}}\ ,\qquad
\vert\Psi_1\rangle=\frac{\vert01\rangle+\vert10\rangle}{\sqrt{2}}\\
\label{Bellstates}
&&
\vert\Psi_2\rangle=\frac{\vert00\rangle-\vert11\rangle}{\sqrt{2}}\ ,\qquad
\vert\Psi_3\rangle=\frac{\vert01\rangle-\vert10\rangle}{\sqrt{2}}\ .
\end{eqnarray}
Under transposition $T[\sigma_\mu]=\varepsilon_\mu\,\sigma_\mu$, $\varepsilon_\mu=(1,1,-1,1)$ and under partial transposition the projection
$P_0$ goes into the flip operator $V\vert\psi\otimes\varphi\rangle=\vert\varphi\otimes\psi\rangle$:
$$
\id\otimes T[P_0]=\frac{1}{2}\,V=\frac{1}{2}\sum_{\mu=0}^3\ v_\mu\, P_\mu\ ,\qquad v_\mu=(1,1,1,-1)\ .
$$
Therefore, the action of partial transposition on a $\sigma$-diagonal state yields
\begin{eqnarray}
\nonumber
\id\otimes T[\rho]&=&\frac{1}{2}\sum_{\mu=0}^3r_\mu\,\1\otimes\sigma_\mu\,V\,\1\otimes\sigma_\mu=
\frac{1}{2}\sum_{\mu,\nu=0}^3r_\mu\,v_\nu\,\1\otimes\sigma_\mu\sigma_\nu\,P_0\,\1\otimes\sigma_\nu\sigma_\mu=
\frac{1}{2}\sum_{\mu,\nu=0}^3 \,v_\nu\,r_\mu\, P_{[\mu,\nu]}\\
\label{PartTransp}
&=&
\frac{1}{2}\sum_{\alpha=0}^3 \Big(\sum_{\nu=0}^3\,v_\nu\,r_{[\alpha,\nu]}\Big)\, P_\alpha=\frac{1}{2}\sum_{\alpha=0}^3 \Big(r_{[\alpha,0]}+r_{[\alpha,1]}+r_{[\alpha,2]}-r_{[\alpha,3]}\Big)\, P_\alpha=\frac{1}{2}\sum_{\alpha=0}^3 \Big(1-2\,r_{[\alpha,3]}\Big)\, P_\alpha\ .
\end{eqnarray}
In the above expression, the following construction has been employed: given $(\alpha,\mu)$, $\alpha,\mu=0,1,2,3$, $[\alpha,\mu]$ is the unique index from $0$ to $3$ such that $\sigma_\alpha\sigma_\mu=\eta^{[\alpha,\mu]}_{\alpha\mu}\,\sigma_{[\alpha,\mu]}$, where
$\eta^{[\alpha,\mu]}_{\alpha\mu}$ is a phase $\pm 1$ or $\pm i$. Because of the Pauli algebraic relations, the symbol $[\cdot,\cdot]$ enjoys the following properties that can be used to derive~(\ref{PartTransp}):
$$
[\alpha,\mu]=[\mu,\alpha]\ ,\qquad [\alpha,\mu]=\gamma\Rightarrow [\alpha,\gamma]=\mu\Rightarrow \ [\mu,\gamma]=\alpha\ .
$$
Thus, $[\alpha,\cdot]$ is a one to one map from the set $(0,1,2,3)$ onto itself.

Since positivity under partial transposition identifies all separable states of two qubits, a $\sigma$-diagonal state is separable if and only if
$\displaystyle r_\mu\leq\frac{1}{2}$ for all $\mu=0,1,2,3$. We now want to recover this result in the light of previous proposition.

From Remark~\ref{rem2}, using~(\ref{Stormer1}),~(\ref{transp2}) and~(\ref{Tr2}), we can write any diagonal positive map $\Lambda_{diag}:M_2(\CI)\mapsto M_2(\CI)$ as
\begin{eqnarray*}
\Lambda_{diag}&=&\sum_{\alpha=0}^3\lambda^{(1)}_\alpha\,S_\alpha\,+\,\sum_{\alpha=0}^3\lambda^{(2)}_\alpha\,S_\alpha\circ T=
\sum_{\alpha=0}^3\lambda^{(1)}_\alpha\,S_\alpha\,+\,\frac{1}{2}\sum_{\alpha,\beta=0}^3\varepsilon_\beta\lambda^{(2)}_\alpha\,S_\alpha\circ S_\beta\\
&=&\sum_{\gamma=0}^3\Big(\lambda^{(1)}_\gamma\,+\,\frac{1}{2}\sum_{\beta=0}^3\varepsilon_\beta\lambda_{[\beta,\gamma]}\Big)\,S_\gamma
=\mu\,\sum_{\gamma=0}^3\
\Big(\frac{1}{2}-\lambda_\gamma\Big)\,S_\gamma\ ,
\end{eqnarray*}
where $\lambda^{(1,2)}_\alpha$ are, according to Theorem~\ref{KS}, positive numbers. Then, the coefficients
$$
\lambda_\gamma=\frac{1}{2}-\frac{1}{\mu}\left(\lambda^{(1)}_\gamma+\frac{1}{2}\sum_{\beta=0}^3\varepsilon_\beta\lambda^{(2)}_{[\beta,\gamma]}\right)
$$
can always be made positive and thus $\Lambda=\sum_{\alpha=0}^3\lambda_\alpha\,S_\alpha$ completely positive, by choosing $\mu$ large enough. Then, they fulfil the condition~(\ref{crit1}) that corresponds to
$\Lambda_{diag}$ being positive. Let us rewrite inequality~(\ref{crit2}) as follows
$$
\frac{1}{2}\,-\,\sum_{\gamma=0}^3\lambda_\gamma\,r_\gamma=\frac{1}{\mu}\sum_{\gamma=0}^3\,r_\gamma\,
\left(\lambda^{(1)}_\gamma+\frac{1}{2}\sum_{\beta=0}^3\varepsilon_\beta\lambda^{(2)}_{[\beta,\gamma]}\right)\,\geq\, 0\ .
$$
Consider the right hand side of the equality; by choosing $\lambda^{(1)}_\gamma=0$ for all $\gamma$ and $\lambda^{(2)}_\gamma=\delta_{\gamma\alpha}$, one gets
$$
\sum_{\beta=0}^3\,r_{[\alpha,\beta]}\,\varepsilon_\beta=r_{[\alpha,0]}+r_{[\alpha,1]}-r_{[\alpha,2]}+r_{[\alpha,3]}=1-2\,r_{[\alpha,2]}\,\geq\, 0
$$
and thus, by varying $\alpha$, $\displaystyle r_\mu\leq\frac{1}{2}$ for all $\mu=0,1,2,3$. Vice versa, if $\displaystyle r_\mu\leq\frac{1}{2}$ for all $\mu=0,1,2,3$, one obtains
\begin{eqnarray*}
\sum_{\gamma=0}^3\,r_\gamma\,\sum_{\beta=0}^3\varepsilon_\beta\lambda^{(2)}_{[\beta,\gamma]}&=&
\sum_{\alpha=0}^3\,\lambda^{(2)}_\alpha\sum_{\beta=0}^3\varepsilon_\beta\,r_{[\alpha,\beta]}=
\sum_{\alpha=0}^3\,\lambda^{(2)}_\alpha\,\Big(r_{[\alpha,0]}+r_{[\alpha,1]}-r_{[\alpha,2]}+r_{[\alpha,3]}\Big)\\
&=&\sum_{\alpha=0}^3\,\lambda^{(2)}_\alpha\,\Big(1-2\,r_{[\alpha,2]}\Big)\geq 0\ ,
\end{eqnarray*}
so that the inequality is satisfied.
\end{example}

\section{Lattice states}

In this section we shall restrict ourselves to the lattice states $\rho_I\in M_{16}(\CI)$ in~(\ref{lattstate}), namely to uniformly distributed $\sigma$-diagonal states with $n=2$ .
Proposition~\ref{propdiagst} now reads

\begin{corollary}
\label{corLS}
A lattice state $\rho_{I}$ is separable if and only if
\begin{equation}
\label{crit2LS}
\sum_{(\alpha,\beta) \in I}\lambda_{\alpha\beta}\,\leq\,\frac{N_{I}}{4}
\end{equation}
for all choices of $16$ coeffcients $\lambda_{\alpha\beta}\geq 0$ such that
\begin{equation}
\label{crit1LS}
\sum_{(\alpha,\beta)\in I}\lambda_{\alpha\beta}\vert\langle\varphi\vert\sigma_{\alpha\beta}\vert\psi\rangle\vert^2\,\leq\,1\qquad\forall\
\vert\varphi\rangle,\vert\psi\rangle\in\CI^4\ .
\end{equation}
Otherwise, if a choice of positive coefficients exists that satisfy~(\ref{crit1LS}) and for which
\begin{equation}
\label{crit3LS}
\sum_{(\alpha,\beta) \in I}\lambda_{\alpha\beta}\,>\,\frac{N_{I}}{4}\ ,
\end{equation}
then a lattice-state $\rho_I$ is entangled.
\end{corollary}

Before drawing concrete conclusions from this result, we examine the entanglement criteria in Propositions~\ref{PPT2} and~\ref{PPT3} in the light of the diagonal structure of witnessing maps which is the main result of the previous section.

\begin{example}
\label{ex3}
The states considered in~(\ref{Ex2}) were found to be entangled by showing that they do not remain positive under the action of $\id\otimes\Gamma_t$, the map
\begin{eqnarray}
\Gamma^t&=&g_{00}(t)\,S_{00}+\sum_{i=1}^{3}\Big(g_{0i}(t)S_{0i}+g_{i0}(t)S_{i0}\Big) \qquad\hbox{with}\\
\nonumber g_{00}(t)&=&\frac{1+3 e^{-4t}}{4}\frac{3+e^{-4t}}{4},\quad g_{0i}(t)=\varepsilon_i\frac{1+3 e^{-4t}}{4}\frac{1-e^{-4t}}{4},\quad g_{i0}(t)=\frac{1- e^{-4t}}{4}\frac{3+e^{-4t}}{4}\ ,
\end{eqnarray}
being proved to be a positive map from $M_4(\CI)$ into itself.
This map, expressed by means of the notation of Example~\ref{ex1} is already in diagonal form; the diagonal completely positive map associated to it by
Proposition~\ref{Stormer} has the form
\begin{eqnarray}
\nonumber
\Lambda^{cp}(t)&=&\Tr-\frac{\Gamma^t}{\mu}\\
\label{aid}
&=&\Big(\frac{1}{4}-\frac{g_{00}(t)}{\mu}\Big)\,S_{00}+\sum_{i=1}^{3}\Big[\Big(\frac{1}{4}-\frac{g_{0i}(t)}{\mu}\Big)\,S_{0i}+
\Big(\frac{1}{4}-\frac{g_{i0}(t)}{\mu}\Big)\,S_{i0}\Big]+\frac{1}{4}\sum_{\alpha,\beta\neq 0}S_{\alpha\beta}\ ,
\end{eqnarray}
with $\mu$ which has to be adjusted taking into account that
$$
g_{0i}(t)\leq\frac{1}{16}\ ,\quad g_{i0}(t)\leq\frac{1}{16}\ ,\quad g_{00}(t)\leq 1\ .
$$
Then, complete positivity of the map $\Lambda^{cp}$ is guaranteed by $\mu \geq\frac{1}{4} $ which yields
$$
\lambda_{00}(t)=\frac{1}{4}-\frac{g_{00}(t)}{\mu}\geq 0\ ,\ \lambda_{0i}(t)=\frac{1}{4}-\frac{g_{0i}(t)}{\mu}\geq 0\ ,\ \lambda_{i0}(t)=\frac{1}{4}-\frac{g_{i0}(t)}{\mu}\geq 0\ ,\ \lambda_{ij}=\frac{1}{4}\ .
$$
These coefficients surely satisfy condition~(\ref{crit1LS}) as the latter just reflects the positivity of the originating map $\Gamma^t$; they also satisfy condition~(\ref{crit3LS}). Indeed,
\begin{eqnarray}
\sum_{(\alpha,\beta)\in I}\lambda_{\alpha\beta}(t)&=&\frac{N_I}{4}-\frac{1}{\mu}\sum_{(\alpha,\beta)\in I}\Big[g_{00}(t))\delta_{\alpha,0}\delta_{\beta,0}+g_{0\beta}(t))\delta_{\alpha,0}
+g_{\alpha 0}(t))\delta_{\beta,0})\Big]\\
\label{gamma2}
&\simeq\atop{t\to 0}&\frac{N_I}{4}-\frac{1}{\mu}\sum_{(\alpha,\beta)\in I}\Big[(1-4t)\delta_{\alpha,0}\delta_{\beta,0}+t(\delta_{\alpha,0}\varepsilon_{\beta}+\delta_{\beta,0})\Big]\ .
\end{eqnarray}
For both subsets in (\ref{Ex2}), the second term in (~\ref{gamma2}) is negative due to $\varepsilon_2=-1$. Thus, $\sum_{(\alpha,\beta)\in I}\lambda_{\alpha\beta}(t)>\frac{N_I}{4}$ for small times.
\end{example}

\begin{example}
\label{ex4}
Let us now consider the lattice state in Example~\ref{Ex3}: in~\cite{benatti2}, it has been shown to be entangled by using the following positive map
\begin{equation}
\label{map3}
M_{4}(\CI)\ni X\mapsto\Phi_V[X]=\Tr[X] - \Big(\rT[X]+\cV[X]\Big) \ ,\quad \cV[X]=V^{\dag}\,X\,V\ ,
\end{equation}
consisting of the trace map to which one subtracts the transposition map and a completely positive map $\cV$ constructed with a $4\times 4$ matrix $V$ such that, in the standard representation,
\begin{equation}
V=\sum_{\alpha\neq2}v_{\alpha2}\sigma_{\alpha2}+\sum_{\beta\neq2}v_{2\beta}\sigma_{2\beta}=-V^T\ ,\quad \sum_{\alpha\neq 2}\Big(|v_{\alpha2}|^2+|v_{2\alpha}|^2\Big)=1\ .
\label{map3.1}
\end{equation}
In this way,
$$
\Phi_V[\vert\psi\rangle\langle\psi\vert]=1-\vert\psi^*\rangle\langle\psi^*\vert-V^\dag\,\vert\psi\rangle\langle\psi\vert\,V=1-P-Q\ ,
$$
where $P$ and $Q$ are orthogonal one-dimensional projections and thus ensure the positivity of the map.

Because of $\cV$, the map $\Phi_V$ is non-diagonal in the maps $S_{\alpha\beta}$: in order to relate the map $\Phi_V$ to Proposition~\ref{propdiagst}, we use~(\ref{transp4}) and get
$$
\Phi^{diag}_V=\sum_{\alpha\neq 2}\Big(\Big(\frac{1}{2}-|v_{\alpha2}|^2\Big)S_{\alpha2}+\Big(\frac{1}{2}-|v_{2\beta}|^2\Big)S_{2\beta}\Big)\ .
$$
The mean value of the Choi matrix of $\Phi^{diag}_V$ with respect to the lattice state in~(\ref{Ex3}) reads
$$
\Tr\Big(\rho_I\,\id\otimes\Phi^{diag}_V[P^4_+]\Big)=\frac{1}{N_I}\Big(\frac{1}{2}-|v_{12}|^2\Big)
$$
and becomes negative choosing $|v_{12}|^2>1/2$ hence revealing the entanglement of $\rho_I$.

Proposition associates to $\Phi^{diag}_V$ completely positive maps of the form
$$
\Lambda^{CP}=\Tr-\frac{\Phi^{diag}_V}{\mu}=\sum_{\alpha\neq 2}\Big(\frac{1}{4}-\frac{1}{2\mu}+\frac{|v_{\alpha2}|^2}{\mu}\Big)S_{\alpha2}+\sum_{\beta\neq 2}\Big(\frac{1}{4}-\frac{1}{2\mu}+\frac{|v_{2\beta}|^2}{\mu}\Big)S_{2\beta}
+\frac{1}{4}\Big(S_{22}+\sum_{\alpha\neq 2,\beta\neq 2}S_{\alpha\beta}\Big)\ ,
$$
whose coefficients are positive if $\mu\geq 2$.
The sum of the coefficients corresponding to the subset $I$ of the lattice state in~(\ref{Ex3}) yields
$$
\sum_{\alpha,\beta\in I}\lambda_{\alpha\beta}=\frac{N_I}{4}-\frac{1}{\mu}\Big(\frac{1}{2}-|v_{12}|^2\Big)\,>\,\frac{N_I}{4}
$$
when $|v_{12}|^2>1/2$ which exposes the entanglement of $\rho_I$, in agreement with~(\ref{crit3LS}).
\end{example}

\subsection{Separable Lattice States}

Because of the convexity of the subset of separable states, one may check whether a lattice state is separable by trying to write it as a convex combination of other lattice states that are known to be separable.
For some $\rho_I$ this is rather straightforward as shown in~\cite{benatti2}; for instance, consider the lattice state
$$
\rho_I=\frac{1}{8}\Big(P_{11}+P_{12}+P_{13}+P_{21}+P_{23}+P_{31}+P_{32}+P_{33}\Big)\ .
$$
According to Proposition~\ref{PPT}, it is PPT; it is also separable, Indeed, the defining subset $I$ splits as follows
$$
\underbrace{\begin{array}{c|c|c|c|c}
               3 & \quad\!  & \times & \times  & \times     \\
               \hline
               2 & \quad\!  &  \times   &  \quad\!  &  \times   \\
               \hline
               1 &  \quad\! & \times &  \times & \times     \\
               \hline
               0 &    &     &     &      \\
               \hline
                 & 0 & 1 & 2 & 3
 \end{array}}_{I}\ =\
 \underbrace{\begin{array}{c|c|c|c|c}
               3 &  \quad\!  &  \times  &  \times   &      \\
               \hline
               2 &  \quad\!  &  \quad\!  & \quad\!  &      \\
               \hline
               1 &  \quad\! & \times &  \times   &       \\
               \hline
               0 & \quad\!  &  \quad\!   &  \quad\! &      \\
               \hline
                 & 0 & 1 & 2 & 3
 \end{array}}_{I_1}\ +\
 \underbrace{\begin{array}{c|c|c|c|c}
               3 &  \quad\!  &  \quad\!  & \quad\!   &      \\
               \hline
               2 &  \quad\! & \times &  \quad\!   & \times     \\
               \hline
               1 &  \quad\! & \times &  \quad\!   & \times     \\
               \hline
               0 &  \quad\!  & \quad\!   &  \quad\!   &      \\
               \hline
                 & 0 & 1 & 2 & 3
 \end{array}}_{I_2}\ +\
 \underbrace{\begin{array}{c|c|c|c|c}
               3 &  \quad\!  &  \quad\!  &  \times &  \times    \\
               \hline
               2 & \quad\!  & \quad\! & \quad\!  &      \\
               \hline
               1 &  \quad\!  &  \quad\!  &  \times   &  \times     \\
               \hline
               0 & \quad\! & \quad\!  & \quad\! &      \\
               \hline
                 & 0 & 1 & 2 & 3
 \end{array}}_{I_3}
\ +\
\underbrace{\begin{array}{c|c|c|c|c}
               3 & \quad\!   & \times   &  \quad\! &  \times    \\
               \hline
               2 & \quad\!  & \times & \quad\!  &  \times    \\
               \hline
               1 &  \quad\!  & \quad\!   & \quad\! &       \\
               \hline
               0 & \quad\! & \quad\!  & \quad\! &      \\
               \hline
                 & 0 & 1 & 2 & 3
 \end{array}}_{I_4}\ .
 $$
The $4$-point subsets $I_i$ are not disjoint, but all points contribute exactly twice to $I$, thence one rewrites
$$
\rho_I=\frac{1}{4}\sum_{i=1}^4\rho_{I_i}\ ,
$$
in terms of rank-$4$ lattice states corresponding to the subsets $I_i$. The result follows since the criterion of Proposition~\ref{PPT} ensures that they are all PPT~\cite{HCLV}.

A more general sufficient condition for the separability of lattice states can be derived by introducing the notion of \textit{special quadruples}.

\begin{definition}
\label{quadruples}
A special quadruple $\QI$ is any subset of the square lattice $L_{16}$ consisting of $4$ points $(\alpha,\beta)$ such that there exist $\vert\varphi\rangle,\vert\psi\rangle\in\CI^4$ for which
\begin{equation}
\label{quaterna0}
\frac{1}{4} \sum_{(\alpha,\beta)\in\QI} \vert\langle\varphi\vert\sigma_{\alpha\beta}\vert\psi\rangle\vert^2 =1\ .
\end{equation}
Given a lattice point $(\alpha,\beta)\in L_{16}$, we will denote by $Q_{\alpha\beta}\in\cQ$ any special quadruple containing $(\alpha,\beta)$, by $\cQ_{\alpha\beta}$ the set of such quadruples and by $n_{\alpha\beta}$ its cardinality.
\end{definition}

We now characterize the set of special quadruples containing $(0,0)\in L_{16}$.

\begin{lemma}
\label{lemmaquadr}
All special quadruples $Q\in\cQ_{00}$ correspond to four commuting $\sigma_{\alpha\beta}$.
\end{lemma}

\proof
Since  $\vert\langle\varphi\vert\sigma_{\alpha\beta}\vert\psi\rangle\vert^2 =1$, a set of $4$ points $\{(\alpha_j,\beta_j)\}_{j=0}^3\subset I$ is
a special quadruple if and only there exist $\vert\psi\rangle,\vert\phi\rangle\in\CI^4$ such that
\begin{equation}
\label{quaterna1}
\sigma_{\alpha_j\beta_j}\vert\psi\rangle={\rm e}^{i\chi_j}\,\vert\phi\rangle\qquad \forall j=0,1,2,3\ .
\end{equation}
Let us focus upon $\cQ_{00}$, the set of all special quadruples $\{(0,0),(\alpha_1,\beta_1),(\alpha_2,\beta_2),(\alpha_3,\beta_3)\}$ containing  the point $(0,0)$.
Each of them  is obtained from the fact that, using~(\ref{quaterna1}) with $(\alpha_0\beta_0)=(00)$,
$$
\vert\varphi\rangle=\vert\psi\rangle\Rightarrow\sigma_{\alpha_j\beta_j}\vert\psi\rangle={\rm e}^{i\chi_j}\,\vert\psi\rangle\ ,\quad j=1,2,3\ .
$$
Therefore,
$$
\sigma_{\alpha_\ell\beta_\ell}\sigma_{\alpha_j\beta_j}\vert\psi\rangle={\rm e}^{i\chi_j}\,\sigma_{\alpha_\ell\beta_\ell}\vert\psi\rangle=
{\rm e}^{i(\chi_j+\chi_\ell)}\,\vert\psi\rangle=\sigma_{\alpha_j\beta_j}\sigma_{\alpha_\ell\beta_\ell}\vert\psi\rangle\ .
$$
Thus, the $\sigma_{\alpha\beta}$ of a special quadruple $Q\in\cQ_{00}$ must commute. Indeed,
\begin{eqnarray}
\nonumber
\Big[\sigma_{\alpha\beta}\,,\,\sigma_{\gamma\delta}\Big]
&=&\sigma_\alpha\sigma_\gamma\otimes\sigma_\beta\sigma_\delta\,-\,
\sigma_\gamma\sigma_\alpha\otimes\sigma_\delta\sigma_\beta=
\Big(1-\epsilon_{\alpha\gamma}\epsilon_{\beta\delta}\Big)\sigma_\alpha\sigma_\gamma\otimes\sigma_\beta\sigma_\delta\\
\label{comrel}
&=&\Big(1-\epsilon_{\alpha\gamma}\epsilon_{\beta\delta}\Big)
\eta^\mu_{\alpha\gamma}\eta^\nu_{\beta\delta}\sigma_{\mu\nu}\ ,
\end{eqnarray}
where $\eta^\mu_{\alpha\beta}$ are the coefficients $\pm1$ and $\pm i$ such that
$\sigma_\alpha\sigma_\beta=\eta^\mu_{\alpha\beta}\sigma_\mu$ and
the $16$ coefficients $\epsilon_{\alpha\gamma}=\pm1$ are given by the Pauli matrix commutation relations.
It thus follows that
$\Big[\sigma_{\alpha\beta}\,,\,\sigma_{\gamma\delta}\Big]\vert\psi\rangle=0$ if and only if
$\epsilon_{\alpha\gamma}\epsilon_{\beta\delta}=1$.
\qed
\medskip

The following properties can be directly checked:
\begin{enumerate}
\item
each $\sigma_{\alpha\beta}$ with $(\alpha,\beta)\neq (0,0)$ commutes with $6$ $\sigma$'s and anti-commute with $8$ $\sigma$'s;
\item
given two commuting $\sigma_{\alpha\beta}\neq\sigma_{00}$ and  $\sigma_{\gamma\delta}\neq\sigma_{00}$ there is a unique $\sigma_{\mu\nu}\neq\sigma_{00}$ commuting  with both of them.
\end{enumerate}
It follows that the set $\cQ_{00}$ consists of the following $15$ special quadruples (omitting the point $(0,0)$ common to all of them)
\begin{eqnarray}
\nonumber
&&\hskip-.5cm
\{(0,1);(1,0);(1,1)\}  \quad     \{(0,2);(2,0);(2,2)\} \quad   \{(0,3);(3,0);(3,3)\}
\quad \{(1,1);(2,2);(3,3)\} \quad \{(1,2);(2,3);(3,1)\}\\
\nonumber
&&\hskip-.5cm
\{(0,1);(2,1);(2,0)\} \quad \{(0,2);(1,2);(1,0)\}  \quad \{(0,3);(1,3);(1,0)\}
\quad \{(1,1);(2,3);(3,2)\} \quad \{(1,3);(2,2);(3,1)\}\\
&&\hskip-.5cm
\{(0,1);(3,1);(3,0)\}   \quad    \{(0,2);(3,2);(3,0)\}  \quad  \{(0,3);(2,3);(2,0)\}
\quad \{(1,2);(2,1);(3,3)\}  \quad \{(1,3);(2,1);(3,2)\}\ .
\label{specQ00}
\end{eqnarray}

The patterns of the special quadruples as well as the proof of the following lemma are reported in Appendix A.
\medskip

\begin{lemma}
\label{lem0}
The special quadruples in the set $\cQ_{00}$ have the following properties:
\begin{enumerate}
\item
two different special quadruples cannot have more than one lattice point in common;
\item
lattice points belonging to different quadruples with a point in common correspond to anti-commuting $\sigma$'s;
\item
each lattice point belongs to three different special quadruples;
\item
let $(\alpha_1,\beta_1)$ and $(\alpha_2,\beta_2)$ be such that $\sigma_{\alpha_1\beta_1}$ and $\sigma_{\alpha_2\beta_2}$
anti-commute and consider the three special quadruples with  $(\alpha_1,\beta_1)$, respectively  $(\alpha_2,\beta_2)$  in common.
They consist of $9$ lattice lattice points different from $(0,0)$,  $(\alpha_1,\beta_1)$ and $(\alpha_2,\beta_2)$ , which divide into three disjoint subsets corresponding to anti-commuting $\sigma$'s.
\end{enumerate}
\end{lemma}
\medskip

Knowledge of $\cQ_{00}$ is sufficient to derive the form of all $\cQ_{\alpha\beta}$; in order to prove this fact, consider the following maps indexed by lattice points $(\alpha,\beta)\in L_{16}$:
\begin{equation}
\label{quadruplemap}
\tau_{\alpha\beta}\,:\,L_{16}\mapsto L_{16}\ ,\qquad \tau_{\alpha\beta}[(\gamma,\delta)]=([\alpha,\gamma],[\beta,\delta])\ ,
\end{equation}
where the map $(\alpha,\gamma)\mapsto[\alpha,\gamma]$ has been introduced in Example~\ref{ex2}. It follows that the maps $\tau_{\alpha\beta}$ are invertible:
$\tau_{\alpha\beta}\circ\tau_{\alpha\beta}[(\gamma,\delta)]=(\gamma,\delta)$. Given a subset $I=\{(\alpha_i,\beta_i)\}\subseteq L_{16}$, $\tau_{\alpha\beta}[I]$ will denote the subset
$\{\tau_{\alpha\beta}[(\alpha_i,\beta_i)]\}$.

\begin{lemma}
\label{lemquadr1}
The map $\cQ_{\alpha\beta}\ni Q\mapsto\tau_{\alpha\beta}[Q]\in\cQ_{00}$ is one-to-one.
\end{lemma}

\proof
If $Q=\{(\alpha_j,\beta_j)\}_{j=0}^3$ is a special quadruple in $\cQ_{\alpha\beta}$, then
$\displaystyle
\sigma_{\alpha_j\beta_j}\vert\psi\rangle={\rm e}^{i\chi_j}\vert\phi\rangle$
for some $\vert\psi\rangle,\vert\phi\rangle\in\CI^4$. Right multiplication by $\sigma_{\alpha\beta}$ yields
$\displaystyle
\sigma_{[\alpha,\alpha_j],[\beta,\beta_j]}\vert\psi\rangle={\rm e}^{i\chi'_j}\sigma_{\alpha\beta}\vert\phi\rangle$.
Then, $\{([\alpha,\alpha_j],[\beta,\beta_j]\}_{j=0}^3=\tau_{\alpha\beta}[Q]$ is a special quadruple for $(0,0)$ exposed by the vectors
$\vert\psi\rangle$ and $\sigma_{\alpha\beta}\vert\phi\rangle$.
The one-to-one correspondence follows from the invertibility of the maps $\tau_{\alpha\beta}$.
\qed
\medskip

\begin{remark}
\label{remquadr}
Local unitary actions of the form $\sigma_{\alpha\beta}\otimes\1\rho_I\sigma_{\alpha\beta}\otimes\1$ transform
$(\gamma,\delta)$ into $\tau_{\alpha\beta}[(\mu,nu)]$ moving $(0,0)$ into $(\alpha,\beta)$. Consider instead rotation matrices  $U,V\in M_2(\CI)$ such that $U\sigma_iU^T=\sigma_k$ and $V\sigma_jV^T=\sigma_\ell$. Local unitary actions as
$$
(U\otimes V)\otimes (U\otimes V)\rho_I(U^T\otimes V^T)\otimes(U^T\otimes V^T)
$$
can be used to exchange the rows $R_j$ and $R_\ell$ and  the columns $C_i$ and $C_k$, while keeping fixed the remaining two rows and columns. Indeed, the state $\vert\Psi^4_+\rangle$ is such that
$$
A\otimes B\vert\Psi^4_+\rangle=\1\otimes (BA^T)\vert\Psi^4_+\rangle\ ,
$$
 where $A^T$ denotes the transposition of  $A$ with respect to  the selected orthonormal basis. Therefore,
\begin{eqnarray*}
&&
\Big((U\otimes V)\otimes (U\otimes V)\Big)\Big((\1\otimes\sigma_{\gamma\delta}\Big)\vert\Psi^4_+\rangle\langle\Psi^4_+\vert\Big(\1\otimes\sigma_{\gamma\delta}\Big) \Big((U^T\otimes V^T)\otimes(U^T\otimes V^T)\Big)=\\
&&\hskip 1cm =\1\otimes(U\sigma_\gamma U^T\otimes V\sigma_\delta V^T)\vert\Psi^4_+\rangle\langle\Psi^4_+\vert
\1\otimes(U^T\sigma_\gamma U\otimes V^T\sigma_\delta V)\ .
\end{eqnarray*}
\end{remark}
\medskip

It is easy to check that all $Q\in\cQ_{00}$ in (\ref{specQ00}) give rise to lattice states $\rho_Q$ that satisfy the criterion in Proposition~\ref{specQ00} for being PPT; this is also true for lattice states corresponding to $Q\in\cQ_{\alpha\beta}$: in fact,
they are obtained from the previous ones by the local action $\sigma_{\alpha\beta}\otimes\1\rho_Q\sigma_{\alpha\beta}\otimes\1$.
The following lemma shows that they are separable as dictated by the general result in~\cite{HCLV}.

\begin{lemma}
\label{lemmaquadr1}
Let $I\subset L_{16}$; all positive coefficients $\lambda_{\alpha\beta}$ satisfying inequality~(\ref{crit2LS}) are such that
$\sum_{(\alpha,\beta)\in Q}\lambda_{\alpha\beta}\leq 1$ for all special quadruples $Q\subseteq I$.
\end{lemma}

\proof
From~(\ref{Tr4}) it follows that
$$
1=\langle\phi\vert\Tr[\vert\psi\rangle\langle\psi\vert]\vert\phi\rangle=\frac{1}{4}\sum_{(\alpha,\beta)\in L_{16}}\left|\langle\phi\vert\sigma_{\alpha\beta}\vert\psi\rangle\right|^2\ .
$$
Therefore, if $\vert \psi\rangle$ and $\vert\phi\rangle$ satisfy~(\ref{quaterna0}) for a given special quadruple $Q$, then $\langle\phi\vert\sigma_{\alpha\beta}\vert\psi\rangle=0$ for all
$(\alpha,\beta)$ not belonging to $Q$. Consider now a set of positive coefficients
$\lambda_{\alpha\beta}$ satisfying
$$
\sum_{(\alpha,\beta)\in I} \lambda_{\alpha\beta}\,\left|\langle\phi\vert\sigma_{\alpha\beta}\vert\psi\rangle\right|^2\leq 1\qquad\forall \vert\psi\rangle,\vert\phi\rangle\in\CI^4\ .
$$
If $Q\subset I$ is a special quadruple and $\vert\psi\rangle$, $\vert\phi\rangle$ satisfy~(\ref{quaterna0}), then
$$
1\geq \sum_{(\alpha,\beta)\in I} \lambda_{\alpha\beta}\,\left|\langle\phi\vert\sigma_{\alpha\beta}\vert\psi\rangle\right|^2=
\sum_{(\alpha,\beta)\in Q} \lambda_{\alpha\beta}\ .
$$
\qed
\medskip

When $N_I=4$, the previous result and Corollary~\ref{corLS} yield

\begin{corollary}
\label{quadruplecor0}
Each rank $4$ lattice state $\rho_I$ with $I\in\cQ$ is separable.
\end{corollary}
\medskip

We now show how special quadruples can be used to detected separable lattice states.
\medskip

\begin{definition}
\label{defcovering}
Given a subset $I\subseteq L_{16}$ we shall term a covering of $I$ any collection $\cQ_I$ of special quadruples (not necessarily disjoint) contained in $I$ such that $\bigcup_{Q\in\cQ_I} Q=I$; we shall denote by $N_{\cQ_I}$ its cardinality.
Further, we shall denote by $M^{\cQ_I}_{\alpha\beta}$ the number of special quadruples in $\cQ_I$ that contains the point $(\alpha,\beta)\in I$
and refer to them as to the multiplicities of $\cQ_I$. Finally, we shall call uniform any covering $\cQ_I$ of $I$ of constant multiplicity, $M^{\cQ_I}_{\alpha\beta}=M_{\cQ_I}$, for all $(\alpha,\beta)\in I$, and minimal if $M_{\cQ_I}$ is the least constant multiplicity.
\end{definition}
\medskip

The usefulness of uniform coverings can be seen as follows: summations over $(\alpha,\beta)\in I$ can be split into sums of contributions from the special quadruples of  any given covering $\cQ_I$ by taking into account to how many
special quadruples $M_{\alpha\beta}^{\cQ_I}$ a point $(\alpha,\beta)$ does belong:
$$
\sum_{(\alpha,\beta)\in I} M_{\alpha\beta}^{\cQ_I}\,\lambda_{\alpha\beta}=\sum_{Q\in\cQ_I}\sum_{(\alpha,\beta)\in Q}\,\lambda_{\alpha\beta}\ .
$$
If the covering is uniform with multiplicity $M_{\cQ_I}$,
$$
\sum_{(\alpha,\beta)\in I}\,\lambda_{\alpha\beta}=\frac{1}{ M_{\cQ_I}}\sum_{Q\in\cQ_I}\sum_{(\alpha,\beta)\in Q}\,\lambda_{\alpha\beta}\ .
$$
\medskip

\begin{lemma}
\label{lem-mincov}
Let $I\subseteq L_{16}$ contain $N_I$ points and $\cQ_I$ be a uniform covering of $I$ of cardinality $N_{\cQ_I}$ and multiplicity $M_{\cQ_I}$; then,
\begin{equation}
\label{unif-mult}
N_{\cQ_I}=\frac{M_{\cQ_I}\, N_I}{4}\ .
\end{equation}
\end{lemma}

\proof
As the covering is uniform, each of the $N_I$ points in $I$ belongs to $M_{\cQ_I}$ special quadruples; on the other hand, the covering consists of $N_{\cQ_I}$ special quadruples each containing $4$ points of $I$.
\qed

Let us now consider a higher rank lattice state corresponding to the subset
\begin{equation}
\label{Ex10}
 \begin{array}{c|c|c|c|c}
               3 & \quad\!  & \times & \times  & \times     \\
               \hline
               2 & \quad\!  &  \times   &  \times  &  \times   \\
               \hline
               1 &  \quad\! & \times &  \times & \times     \\
               \hline
               0 & \times   &     &     &      \\
               \hline
                 & 0 & 1 & 2 & 3
 \end{array}  \qquad .
 \end{equation}
Let us focus upon $(0,0)\in L_{16}$; the special quadruples in $\cQ_{00}$, given by the last two columns in~(\ref{specQ00}), corresponding to the patterns (\ref{mincov0}) e (\ref{mincov1}) cover $I$, but not uniformly for the multiplicities are
$M^{\cQ_I}_{00}=6$ and $M^{\cQ_I}_{\alpha\beta}=2$ for all other points in $I$.
A uniform covering of $I$ is provided by the first special quadruple in~(\ref{mincov0}) and the second one in~(\ref{mincov1})
$$
\begin{array}{c|c|c|c|c}
               3 &  \quad\!  &  \quad\!  &  \quad\!   &  \times    \\
               \hline
               2 &  \quad\!  &  \quad\!  & \times  &      \\
               \hline
               1 &  \quad\! & \times &  \quad\!   &       \\
               \hline
               0 & \times  &  \quad\!   &  \quad\! &      \\
               \hline
                 & 0 & 1 & 2 & 3
 \end{array}
\quad,\quad
 \begin{array}{c|c|c|c|c}
               3 &  \quad\!  &  \times  & \quad\!   &      \\
               \hline
               2 &  \quad\! & \quad\! &  \quad\!   & \times     \\
               \hline
               1 &  \quad\! & \quad\! &  \times   & \quad\!    \\
               \hline
               0 &  \times  & \quad\!   &  \quad\!   &      \\
               \hline
                 & 0 & 1 & 2 & 3
 \end{array}\quad ,
$$
plus other three ones coming from the first one in~(\ref{mincovrect1}) by using suitable local unitary operations as explained in Lemma~\ref{lemquadr1} :
$$
\begin{array}{c|c|c|c|c}
               3 &  \quad\!  &  \quad\!  &  \times &  \times    \\
               \hline
               2 & \quad\!  & \quad\! & \quad\!  &      \\
               \hline
               1 &  \quad\!  &  \quad\!  &  \times   &  \times     \\
               \hline
               0 & \quad\! & \quad\!  & \quad\! &      \\
               \hline
                 & 0 & 1 & 2 & 3
 \end{array}\quad ,\quad
\begin{array}{c|c|c|c|c}
               3 & \quad\!   & \quad\!   &  \quad\! &  \quad\!    \\
               \hline
               2 & \quad\!  & \times & \quad\!  &  \times    \\
               \hline
               1 &  \quad\!  & \times   & \quad\! &  \times     \\
               \hline
               0 & \quad\! & \quad\!  & \quad\! &      \\
               \hline
                 & 0 & 1 & 2 & 3
 \end{array}\quad ,\quad
\begin{array}{c|c|c|c|c}
               3 & \quad\!   & \times   &  \times &  \quad\!    \\
               \hline
               2 & \quad\!  & \times & \times  &  \quad\!    \\
               \hline
               1 &  \quad\!  & \quad\!   & \quad\! &  \quad\!     \\
               \hline
               0 & \quad\! & \quad\!  & \quad\! &      \\
               \hline
                 & 0 & 1 & 2 & 3
 \end{array}\quad .
$$
This is a uniform covering with multiplicity $M_{\cQ_I}=2$; it is also a minimal covering as,
by~(\ref{unif-mult}), $4 N_{\cQ_I}=10\times M_{\cQ_I}$ implies $M_{\cQ_I}$ even.

Then, the lattice state corresponding to the subset~(\ref{Ex10}) can be convexly decomposed in terms of PPT separable rank-$4$ lattice states:
$$
\rho_I=\frac{1}{10}\Big(P_{00}+P_{11}+P_{12}+P_{13}+P_{21}+P_{22}+P_{23}+P_{31}+P_{32}+P_{33}\Big)=\frac{1}{5}\sum_{j=1}^5\rho_{Q_j}\ ,
$$
where $\{Q_j\}_{j=1}^5$ are the special quadruples of the minimal covering and is thus separable.

The previous argument can be generalized as follows.
\medskip

\begin{proposition}
 \label{theo3}
Suppose $I\subseteq L_{16}$ is a subset with uniform covering $\cQ_I$ of cardinality $N_{\cQ_I}$ and multiplicity $M_{\cQ_I}$, then $\rho_I$ can be convexly decomposed as
\begin{equation}
\label{cov-dec-ls}
\rho_I=\frac{1}{N_{\cQ_I}}\sum_{j=1}^{N_{Q_j}}\rho_{Q_j}\ ,
\end{equation}
where $Q_j\subseteq I$ are the special quadruples in $\cQ$,
and is thus separable.
\end{proposition}

\proof
Let $Q_j$, $1\leq i\leq N_{\cQ_I}$, be the elements of the uniform covering $\cQ_I$. From Lemma~\ref{lemmaquadr1} it follows that
$$
N_{\cQ_I}\geq\sum_{j=1}^{N_{\cQ_I}}\sum_{(\alpha,\beta)\in Q_j} \lambda_{\alpha\beta}=M_{\cQ_I}\,\sum_{(\alpha,\beta)\in I}\lambda_{\alpha\beta}\
\Longrightarrow\
\sum_{(\alpha,\beta)\in I}\lambda_{\alpha\beta}\,\leq\,\frac{N_{\cQ_I}}{M_{\cQ_I}}=\frac{N_I}{4}
$$
so that Corollary~\ref{corLS} ensures separability. Furthermore, using~(\ref{unif-mult}),
$$
\rho_I=\frac{1}{N_I}\sum_{(\alpha,\beta)\in I}P_{\alpha\beta}=\frac{4}{M_{\cQ_I}\,N_{\cQ_I}}\sum_{j=1}^{N_{\cQ_I}}\,\frac{1}{4}\sum_{(\alpha,\beta)\in Q_j}P_{\alpha\beta}=\frac{1}{N_{\cQ_I}}\sum_{j=1}^{\cQ_I}\rho_{Q_j}\ .
$$
\qed
\medskip

Appendix B shows other instances of lattice states whose separability which could not be previously ascertained now follows because of the uniform covering argument.

Opposite to subsets with uniform coverings that lead to separable states, stand those subsets where at least one point does not belong to any special quadruple contained in $I$.These states will be studied in the next section.

\subsection{Entanglement and lattice geometry}

In this section we single out geometric patterns of subsets $I\subseteq L_{16}$ such that $\rho_I$ is surely entangled.

With reference to Corollary~\ref{corLS}, given a subset $I\subseteq L_{16}$, choose the coefficients $\lambda_{\alpha\beta}\geq 0$
such that for a certain $(\alpha_0,\beta_0)\in I$
\begin{equation}
\label{ent-suff-cond}
\lambda_{\alpha_{0}\beta_{0}}=\frac{1+\delta}{4}\ , \qquad \lambda_{\alpha\beta}=\frac{1}{4} \quad \forall (\alpha,\beta)\neq (\alpha_{0},\beta_{0})\ ,
\end{equation}
where $\delta> 0$ is a suitable parameter. Then,
$\displaystyle
\sum_{(\alpha\beta)\in I} \lambda_{\alpha\beta} = \frac{N_{I}+\delta}{4}>\frac{N_{I}}{4}
$
and inequality~(\ref{crit3LS}) is satisfied.
According to the same corollary, in order to conclude that the corresponding lattice state $\rho_I$ is entangled, also inequality~(\ref{crit1LS}) must be fulfilled; namely, one must find $\delta>0$ such that
\begin{equation}
\label{pos-witn}
\frac{\delta}{4}{\vert\langle\varphi\vert\sigma_{\alpha_{0}\beta_{0}}\vert\psi\rangle\vert}^2+
\frac{1}{4}\sum_{(\alpha,\beta)\in I}{\vert\langle\varphi\vert\sigma_{\alpha\beta}\vert\psi\rangle\vert}^2\leq1 ,\quad \forall\vert\varphi\rangle,\vert\psi\rangle\in \CI^{4}\ .
\end{equation}
\medskip

If $(\alpha_0,\beta_0)$ belongs to a special quadruple contained in $I$, such a $\delta>0$ cannot exist; indeed, if $\vert\psi\rangle$ and $\vert\varphi\rangle$ satisfy~(\ref{quaterna0}), then
$\vert\langle\varphi\vert\sigma_{\alpha_{0}\beta_{0}}\vert\psi\rangle\vert>0$ and
$\displaystyle \frac{1}{4}\sum_{(\alpha,\beta)\in I}{\vert\langle\varphi\vert\sigma_{\alpha\beta}\vert\psi\rangle\vert}^2=1$, together with inequality~(\ref{pos-witn}) yield $\delta=0$.
\medskip

\begin{remark}
\label{rem2a}
Suppose that inequality~(\ref{pos-witn}) can be satisfied by a fixed $\delta>0$ fro all $\vert\psi\rangle$ and $\vert\varphi\rangle$. Then, using~(\ref{Tr4}) and~(\ref{Stormer1}), the choice of coefficients in~(\ref{ent-suff-cond}) corresponds to a witness of the form
\begin{equation}
\label{ent-witn}
\Lambda=\Tr-\Lambda_{cp}=\frac{1}{4}\sum_{(\alpha,\beta)\notin I}S_{\alpha\beta}\,-\,\frac{\delta}{4}\,S_{\alpha_0\beta_0}\ .
\end{equation}
By comparison with the transposition in~(\ref{transp4}), one sees that, unlike the $6$ negative contributions in the latter, $\Lambda$ presents only one negative contribution. Yet, we will show that for a family of special subset $\cI$, corresponding
to PPT states, $\Lambda$ provides an entanglement witness.
\end{remark}
\medskip

We now show that lattice states $\rho_I$ for which at leat one point of $I$ does not belong to special quadruples contained in $I$, such a $\delta>0$ fulfilling~(\ref{pos-witn}) can indeed be found whence such states are entangled.
\medskip

\begin{theorem}
\label{theo2}
Given a lattice state $\rho_{I}$, if there exists a point $(\alpha_{0},\beta_{0})\in I$ such that
$\QI_{\alpha_{0}\beta_{0}}\nsubseteq I$ for all $\QI_{\alpha_{0}\beta_{0}}\in\cQ$, then $\rho_I$ is entangled.
\end{theorem}

\noindent
\proof
According to Lemma~\ref{lemquadr1}, the point $(\alpha_0,\beta_0)$ can be transformed into $(0,0)$ and $I$ into a new set, that we shall denote again by $I$ for sake of simplicity, without altering the entanglement or separability of the transformed $\rho_I$ with respect to the initial one.
Then, the assumption of the theorem translates into the fact that no special quadruple in the list~(\ref{specQ00}) is contained in $I$.
Having set  $(\alpha_0,\beta_0)=(0,0)$, inequality~(\ref{pos-witn}) now reads
\begin{equation}
\label{pos-witn1}
\Delta^{\psi,\varphi}_{I,\delta}=\frac{\delta}{4}{\vert\langle\varphi\vert\psi\rangle\vert}^2+
\Delta^{\psi,\varphi}_I\ ,\quad\hbox{where}\quad
\Delta^{\psi,\varphi}_I=\frac{1}{4}\sum_{(\alpha\beta)\in I}{\vert\langle\varphi\vert\sigma_{\alpha\beta}\vert\psi\rangle\vert}^2\ .
\end{equation}
It proves convenient to introduce the following $\psi$-dependent $4\times 4$ matrices
\begin{equation}
\label{pos-witn2}
\hD^\psi_{I,\delta}=\frac{\delta}{4}\vert\psi\rangle\langle\psi\vert+\hD^\psi_I\ ,\quad \hD^\psi_I=\frac{1}{4}\sum_{(\alpha\beta)\in I}\,\sigma_{\alpha\beta}\vert\psi\rangle\langle\psi\vert\sigma_{\alpha\beta}\ ,
\end{equation}
so that one has to prove that
\begin{equation}
\label{pos-witn3}
\exists \delta>0\quad\hbox{such that} \quad \Delta^{\psi,\varphi}_{I,\delta}=\langle\varphi\vert\hD^\psi_{I,\delta}\vert\varphi\rangle\leq 1\quad
\forall\ \vert\psi\rangle,\,\vert\varphi\rangle\in\CI^4\ .
\end{equation}
The major obstruction to $\Delta^{\psi,\varphi}_{I,\delta}\leq1$ with $\delta>0$ arises when $\hD^\psi_I$ has eigenvalue $1$. Since, from~(\ref{Tr4}),
\begin{equation}
\label{pos-witn4}
\hD^\psi_I+\hD^\psi_{I^c}=\1\Longrightarrow 0\leq \hD^\psi_I\leq \1\ ,
\end{equation}
where $I^c$ denotes the complement $L_{16}\setminus I$, the corresponding eigenvectors, $\hD^{\psi}_{I}\vert\varphi\rangle=\vert\varphi\rangle$, satisfy
\begin{equation}
\label{D-comp}
\langle\varphi\vert\hD^{\psi}_{I^{c}}\vert\varphi\rangle=
\frac{1}{4}\sum_{(\alpha,\beta)\in I^c}|\langle\varphi\vert\sigma_{\alpha\beta}\vert\psi\rangle|^2=0
\Longleftrightarrow \vert\varphi\rangle\perp\sigma_{\alpha\beta}\vert\psi\rangle\quad\forall\ (\alpha,\beta)
\in I^c\ .
\end{equation}
Let the eigenvalues of $\hD^\psi_I$ be decreasingly ordered and consider the spectral decomposition
\begin{equation}
\label{pos-witn5}
\hD^\psi_I=P^\psi_I(1)+R^\psi_I\ ,\quad R^\psi_I=\sum_{j>1}d^\psi_I(j)\,P^\psi_I(j)\ ,
\end{equation}
where $P^\psi_I(1)$ projects onto the eigenspace relative to the eigenvalue $1$ and $P^\psi_I(j)$ are the other orthogonal spectral projections relative to the eigenvalues $0\leq d^\psi_I(j)<1$.
Then, if $P^\psi_I(1)\neq 0$, $\delta>0$ in~(\ref{pos-witn3}) is only possible with $P^\psi_I(1)\vert\psi\rangle=0$. These preliminary considerations allow us to prove~(\ref{pos-witn}) through a series of lemmas and corollaries.
\qed
\medskip

\begin{lemma}
\label{lem1}
With the notation of~(\ref{pos-witn5}), if $P^\psi_I(1)\vert\psi\rangle=0$ for all $\vert\psi\rangle\in\CI^4$, then~(\ref{pos-witn3}) is satisfied.
\end{lemma}
\medskip

\noindent
\proof
If $\displaystyle M_I=\sup_{\vert\psi\rangle\in\CI^4}\|R^\psi_I\|=1$, then, by compactness, there exists a converging sequence $\psi_n\to\psi^*$ of vectors in $\CI^4$ such that $R^{\psi_n}_I\perp P^{\psi_n}_I(1)$ and $\|R^{\psi_n}_I\|\to 1$, hence $\hD^{\psi_n}_I$ converges in norm to $\hD^{\psi^*}_I$ with $\|R^{\psi^*}_I\|=1$  which is a contradiction. Therefore, $M_I<1$; hence, choosing $0<\delta\leq 4(1-M_I)$, from $P^\psi_I(1)\vert\psi\rangle=0$ and $P^\psi_I(1)R^\psi_I=0$, one gets
$$
\|\hD^\psi_{I,\delta}\|=
\max\left\{1,\left\|\frac{\delta}{4}\vert\psi\rangle\langle\psi\vert+R^\psi_I\right\|\right\}\leq
\max\left\{1,\frac{\delta}{4}+M_I\right\}\leq 1\ .
$$
\qed
\medskip

\begin{corollary}
\label{cor-ort}
Given a lattice state $\rho_I$ and $\vert\psi\rangle\in\CI^4$, let $V_{I^c}$ be the subspace spanned by
the vectors $\sigma_{\alpha\beta}\vert\psi\rangle$ with $(\alpha,\beta)\in I^c$. If $V^\psi_{I^c}=\CI^4$ for all $\vert\psi\rangle\in\CI^4$, then $\rho_I$ is entangled.
\end{corollary}
\medskip

\proof
From~(\ref{D-comp}) it follows that $P^\psi_I(1)=0$ for all $\vert\psi\rangle\in\CI^4$; thus Lemma~\ref{lem1} applies.
\qed
\medskip

Based on the previous two results, we now focus upon when $P^\psi_I(1)\neq 0$ and show that it projects onto a subspace orthogonal to $\vert\psi\rangle$.

\begin{lemma}
\label{lem2}
If $\sigma_{\mu\nu}\vert\psi\rangle=\pm\vert\psi\rangle$ for some $(\mu,\nu)\in L_{16}$, then, with the notation of~(\ref{pos-witn5}), $P^\psi_I(1)\vert\psi\rangle=0$.
\end{lemma}

\proof
If $(\mu,\nu)\in I^{c}$ and $\langle\varphi\vert\hD^{\psi}_{I}\vert\varphi\rangle=1$,
then, from~(\ref{D-comp}), $\vert\varphi\rangle\perp\sigma_{\alpha\beta}\vert\psi\rangle$ for all $(\alpha,\beta)\in I^{c}$, hence to $\pm\vert\psi\rangle=\sigma_{\mu\nu}\vert\psi\rangle$.
Suppose then that $(\mu,\nu)\in I$ and rewrite inequality~(\ref{pos-witn3}) as
$$
\frac{\delta}{4}{\vert\langle\varphi\vert\psi\rangle\vert}^{2}+\frac{1}{4}\sum_{(\alpha,\beta)\in I_{1}}{\vert\langle\varphi\vert\sigma_{\alpha\beta}\vert\psi\rangle\vert}^{2}+
\frac{1}{4}\sum_{(\alpha,\beta)\in I_{2}}{\vert\langle\varphi\vert\sigma_{\alpha\beta}\vert\psi\rangle\vert}^{2}
\leq 1\ ,
$$
where the index set $I$ has been split into
$$
I_{1}=\{(\alpha,\beta)\in I: [\sigma_{\alpha\beta},\sigma_{\mu\nu}]=0\}\quad\hbox{and}\quad
I_{2}=\{(\alpha,\beta)\in I: \{\sigma_{\alpha\beta},\sigma_{\mu\nu}\}=0\}\ .
$$
The vectors $\sigma_{\alpha\beta}\vert\psi\rangle$ from these two subsets are orthogonal; indeed,
$[\sigma_{\gamma\delta},\sigma_{\mu\nu}]=0$ and $\{\sigma_{\alpha\beta},\sigma_{\mu\nu}\}=0$ yield
$$
\langle\psi\vert\sigma_{\alpha\beta}\sigma_{\gamma\delta}\vert\psi\rangle=
\langle\psi\vert\sigma_{\mu\nu}\sigma_{\alpha\beta}\sigma_{\gamma\delta}
\sigma_{\mu\nu}\vert\psi\rangle
=-\langle\psi\vert\sigma_{\alpha\beta}\sigma_{\gamma\delta}\vert\psi\rangle=0\ .
$$
Therefore, the following two ones are orthogonal matrices:
\begin{eqnarray}
\label{ort1}
\hD^{\psi}_{I_{1}}&=&\frac{1}{2}\vert\psi\rangle\langle\psi\vert
+\frac{1}{4}\sum_{I_1\ni(\alpha,\beta)\neq(00),(\mu,\nu)}
\sigma_{\alpha\beta}\vert\psi\rangle\langle\psi\vert\sigma_{\alpha\beta}\\
\label{ort2}
\hD^{\psi}_{I_{2}}&=&\frac{1}{4}\sum_{(\alpha,\beta)\in I_2}
\sigma_{\alpha\beta}\vert\psi\rangle\langle\psi\vert\sigma_{\alpha\beta}\ .
\end{eqnarray}
It turns out that  $\|\hD^{\psi}_{I_{1}}\|<1$ whence $\hD^\psi_I\vert\varphi\rangle=\vert\varphi\rangle$ can only be due to $\hD^{\psi}_{I_{2}}\vert\varphi\rangle=\vert\varphi\rangle$ and $\vert\varphi\rangle\perp\vert\psi\rangle$.

The fact that  $\|\hD^{\psi}_{I_{1}}\|<1$ can be seen as follows: at most three $\sigma_{\alpha\beta}$ may contribute to the sum in $\hD^{\psi}_{I_{1}}$ and they must anti-commute. In fact, if there were two commuting $\sigma_{\alpha\beta}$ contributing to the sum, they would commute with $\sigma_{\mu\nu}$ and according to Lemma~\ref{lem0}, they would form a special quadruple $Q_{00}$ contained in $I$ which is excluded by hypothesis. Therefore, the $\sigma_{\alpha\beta}$ contributing to the sum must anti-commute. From Lemma~\ref{lem0} again, there cannot be  more than three.
Suppose this is the case; denote by $S_\alpha$, $\alpha=1,2,3$, these three $\sigma_{\alpha\beta}$ such that $\{\sigma_{\alpha\beta},\sigma_{\mu\nu}\}=0$ and rewrite
$$
\hD^{\psi}_{I_{1}}=\frac{1}{4}\vert\psi\rangle\langle\psi\vert
+\frac{1}{4}\sum_{\alpha=0}^3S_\alpha\vert\psi\rangle\langle\psi\vert S_\alpha\ ,\quad S_0=\1_4\ .
$$
Without restriction, we choose $\vert\psi\rangle$ such that $\sigma_{\mu\nu}\vert\psi\rangle=\vert\psi\rangle$.
Each $S_\alpha\vert\psi\rangle$ is an eigenstate of $\sigma_{\mu\nu}$ belonging to the same twice degenerate eigenvalue $-1$. Let $P$ project onto the corresponding eigenspace; then, $[S_\alpha,P]=0$ and the rank $2$ matrices $T_\alpha=P\,S_\alpha\,P=S_\alpha P=P\,S_\alpha$ satisfy the Pauli algebra with identity $P$. Thus,~(\ref{Tr2}) holds with $T_\alpha$ replacing $\sigma_\alpha$, yielding
$$
\hD^{\psi}_{I_{1}}=\frac{1}{4}\vert\psi\rangle\langle\psi\vert
+\frac{1}{4}\sum_{\alpha=0}^3\,S_\alpha\vert\psi\rangle\langle\psi\vert S_\alpha=
\frac{1}{4}\vert\psi\rangle\langle\psi\vert+\frac{1}{2}P\ .
$$
If there are less than three anti-commuting contributions $S_\alpha$, $\alpha\neq 0$, then the second equality becomes a strict inequality. Therefore, $\|\hD^\psi_{I_1}\|\leq3/4<1$.
\qed
\medskip

Let us now consider the set of $\sigma_{\alpha\beta}$ indexed by the points $(\alpha,\beta)$ in the complement set $I^c$: we list them as $(\alpha_i,\beta_i)$, $1\leq i\leq N-N_I$.
The following observations follow from Lemma~\ref{lem0} and concern the ways in which the points of $I^c$ eliminate the special quadruples in $\cQ_{00}$ from being contained in $I$.
\begin{enumerate}
\item
Since $(\alpha_1,\beta_1)\notin I$, the three quadruples it belongs to according to Lemma \ref{lem0} cannot be entirely contained in $I$.
\item
If $(\alpha_2,\beta_2)\in I^c$ does not belong to any of the three quadruples containing $(\alpha_1,\beta_1)$, then three more special quadruples in~(\ref{specQ00}) are not contained in $I$. Moreover, $\sigma_{\alpha_1\beta_1}$ anti-commutes with $\sigma_{\alpha_2\beta_2}$.
It thus follows that the minimum cardinality of $I^c$ complying with the hypothesis of Theorem~\ref{theo2}, namely that no special quadruple of $(0,0)$ is contained in $I$, is $5$. The corresponding $\sigma_{\alpha\beta}$ form an anti-commuting set.
\item
If $(\alpha_2,\beta_2)$ belongs to one special quadruple containing $(\alpha_1,\beta_1)$, then only $2$
new special quadruples are not contained in $I$.
\item
If $(\alpha_3,\beta_3)$ belongs to one of the special quadruples containing $(\alpha_1,\beta_1)$ and one of those containing $(\alpha_2,\beta_2)$, then only $1$ new special quadruple adds to the list of those not contained in $I$.
\item
If $(\alpha_4,\beta_4)$ belongs to three quadruples in the list of those not contained in $I$, it does not add any new quadruple to the list.
\end{enumerate}

\begin{proposition}
\label{lastprop}
If $I\subseteq L_{16}$ is such that it does not contain any special quadruple in $\cQ_{00}$, its complement must surely contain at least three points
$(\alpha_i,\beta_i)$, $i=1,2,3$, such that the corresponding $\sigma_{\alpha_i,\beta_i}$ anti-commute.
\end{proposition}

\proof
First we notice that given the hypothesis and the properties $1-5$ above, $I^c$ must contain at least two points $(\alpha_1,\beta_1)$, $(\alpha_2,\beta_2)$  such that $\sigma_{\alpha_1\beta_1}$ and $\sigma_{\alpha_2\beta_2}$ anti-commute: these two points eliminate $6$ special quadruples from those possibly contained in $I$.\\
We now argue by contradiction and suppose that $I^c$ cannot give rise to three anti-commuting $\sigma$'s. It follows that in order to avoid the remaining $9$ special quadruples to be contained in $I$, the points in $I^c$ must belong to the $3$ special quadruples
of the form~(\ref{specQ00}) containing $(\alpha_1,\beta_1)$ or to the $3$ ones containing $(\alpha_2,\beta_2)$. Indeed, then the corresponding $\sigma$'s commute with at least one of $\sigma_{\alpha_1\beta_1}$ and $\sigma_{\alpha_2\beta_2}$ and cannot constitute a set of three anti-commuting $\sigma$'s.
From the last property of Lemma~\ref{lem0}, these special quadruples contain $3$ disjoint sets of points corresponding to $3$ sets of anti-commuting $\sigma$'s; therefore, since we assumed these sets cannot arise from $I^c$, at least one point from each of them cannot belong to $I^c$, whence there are only $9-3=6$ points to choose from in order to avoid the remaining $9$ special quadruples to be contained in $I$. This can be achieved only if there are at least two points $(\alpha_3,\beta_3)$ and $(\alpha_4,\beta_4)$ among them corresponding to anti-commuting $\sigma_{\alpha_3\beta_3}$ and $\sigma_{\alpha_4\beta_4}$. Because they commute with at least one of $\sigma_{\alpha_1\beta_1}$ and of $\sigma_{\alpha_2\beta_2}$, they eliminate at most $4$ special quadruples from those possibly contained in $I$; so, at least $5$ $Q\in\cQ_{00}$ could still be contained in $I$.
Since the $\sigma$'s corresponding to the remaining $4$ points in $I^c$ we can choose from must commute with at least one of  $\sigma_{\alpha_1\beta_1}$ and $\sigma_{\alpha_2\beta_2}$ and one of
$\sigma_{\alpha_3\beta_3}$ and $\sigma_{\alpha_4\beta_4}$, they cannot eliminate more than $4$ special quadruples from those contained in $I$.
\qed
\medskip

\begin{corollary}
\label{lastcor}
Given a subset $I\subseteq L_{16}$, in order that no special quadruple of $\cQ_{00}$ be contained in $I$, the complement $I^c$ must have the following structure
\begin{enumerate}
\item
$I^c$ must contain  at least $5$ points;
\item
$\hbox{card}(I^c)=5$, then the matrices $\{\sigma_{\alpha_i\beta_i}\}_{i=1}^5$, with $(\alpha_i,\beta_i)\in I^c$ anti-commute;
\item
if $\hbox{card}(I^c)>5$ and there are no sets of five anticommuting $\sigma$'s, then there must exist at least one set consisting of three  $\sigma$'s that anti-commute among themselves plus a fourth one that commutes with only one of them.
\end{enumerate}
\end{corollary}
\medskip

\proof
Since each point can prevent at most $3$ special from being contained in $I$, then $card(I^c)\geq 5$.\\
If $5$ points in $I^c$ correspond to a set of $5$ anti-commuting $\sigma$'s, then, according to Lemma~\ref{lem0}, each of these $5$ points corresponds to $3$ different special quadruples: all of them are thus prevented from being contained in $I$.\\
Suppose the sets of anti-commuting $\sigma$'s that can arise from $I^c$ are of cardinality at most $4$.
From the previous proposition, there is at least one set of $3$ points corresponding to anti-commuting $\sigma$'s: these points prevent $9$ special quadruples in $\cQ_{00}$ from being contained in $I$. Suppose all remaining points in $I^c$  correspond to $\sigma$'s that commute with at least $2$ $\sigma$'s in the anti-commuting set; each of these points can prevent at most $1$ special quadruple from being contained in $I$ and does so only if the corresponding $\sigma$'s do not commute with any other in their set. Since we need $6$ of them, this cannot happen because, by hypothesis, we can have at most sets of $4$ anti-commuting $\sigma$'s.
\qed
\medskip

The next two lemmas concern case $2$ in the previous corollary, where $I^c$ corresponds to a set of $5$ anti-commuting $\sigma_{\alpha\beta}$: the first lemma specifies the structure of the anti-commuting set, while the second one is about the sub-space generated by these matrices acting on a vector state $\vert\psi\rangle$.
\medskip

\begin{lemma}
\label{5anticom}
All sets of $5$ anti-commuting $\sigma_{\alpha_\ell\beta_\ell}$, $(\alpha_\ell,\beta_\ell)\in I^c$, must be of the form
$$
K_1=\{\sigma_{0i}\,,\,\sigma_{0j}\,,\,\sigma_{1k}\,,\,\sigma_{2k}\,,\,\sigma_{3k}\}\quad\hbox{or}\quad
K_2=\{\sigma_{i0}\,,\,\sigma_{j0}\,,\,\sigma_{k1}\,,\,\sigma_{k2}\,,\,\sigma_{k3}\}
$$
with $i,j,k$ such that the anti-symmetric tensor $\varepsilon_{ijk}\neq 0$.
\end{lemma}
\medskip

\proof
Since we have $5$ indices $\alpha_\ell$ and $\beta_\ell$ to choose among $0,1,2,3$, at least two $\alpha$ s
and two $\beta$ s must be equal: let us consider $\alpha_1=\alpha_2$.

If $\alpha_1=\alpha_2=0$, in order to have $\{\sigma_{0\beta_1}\,,\,\sigma_{0\beta_2}\}=0$, the corresponding $\beta_1$ and $\beta_2$ must be different and different from $0$; therefore, $\beta_1=i$, $\beta_2=j$, with $i\neq j$, $i\neq 0$, $j\neq 0$, so that the first two anti-commuting matrices are $\sigma_{0i}$ and $\sigma_{0j}$.
In the three $(\alpha_\ell\beta_\ell)$ left, there cannot appear $\alpha_\ell=0$, otherwise the corresponding $\beta_\ell$, which cannot be $0$ and must equal either $i$ or $j$, would lead to a $\sigma_{\alpha_\ell\beta_\ell}$ violating anti-commutativity. Then, with the three remaining $\alpha_\ell\neq 0$, the three corresponding $\beta_\ell$ must be different from $0$,  $i$ and $j$, otherwise $\sigma_{\alpha_\ell\beta_\ell}$ would commute with either $\sigma_{0i}$ or $\sigma_{0j}$. Thus, $\beta_3=\beta_4=\beta_5=\beta_k$ with $k\neq i$, $k\neq j$, while the corresponding $\alpha_\ell$ must be different, namely $\alpha_3=1$, $\alpha_4=2$ and $\alpha_5=3$. This gives the set $K_1$

If $\alpha_1=\alpha_2=k\neq 0$, the corresponding $\beta_{1,2}$ cannot be equal and must be both different from $0$. Thus, two $\sigma_{\alpha_\ell\beta_\ell}$ in the anti-commuting set are $\sigma_{ki}$ and
$\sigma_{kj}$, with $i\neq j=1,2,3$. Only one of the remaining $\alpha_\ell$ can be $0$, otherwise we would be back to the previous case: for $\sigma_{0\beta_\ell}$ to anti-commute with $\sigma_{ki}$ and $\sigma_{kj}$, $\beta_\ell$ must be equal to $s$ with $s\neq i$, $s\neq j$.
In order to anti-commute with $\sigma_{ki}$, $\sigma_{kj}$ and $\sigma_{0s}$, $\sigma_{\alpha_4\beta_4}$ and
$\sigma_{\alpha_5\beta_5}$, with $\alpha_{4,5}\neq 0$, must be of the form $\sigma_{pi}$ and $\sigma_{qj}$ with $p\neq q$ and both different from $0$ and $k$. But then, they would commute among themselves; thus, $\alpha_{3,4,5}\neq 0$. At least one of them, say $\alpha_3$ must equal $k$; then, anti-commutativity only holds if $\beta_3=r\neq 0$ with
$i\neq r$, $j\neq r$. Finally, since at least two $\beta$'s must be equal, in order to anti-commute among themselves and with $\sigma_{ki}$, $\sigma_{kj}$ and
$\sigma_{kr}$, $\sigma_{\alpha_4\beta_4}$ and $\sigma_{\alpha_5\beta_5}$ must have $\beta_4=\beta_5=0$ and $\alpha_4=m$, $\alpha_5=n$ such that the antisymmetric tensor $\epsilon_{kmn}\neq 0$.
This fixes the set $K_2$. The same result would obtain if one started
arguing with two equal $\beta$'s instead of $\alpha$'s.
\qed
\medskip

\begin{lemma}
\label{5lin-ind}
Let $K_{1,2}$ be the two sets of $5$ anti-commuting $\sigma_{\alpha\beta}$; for any $\vert\psi\rangle\in\CI^4$, let
$$
K^\psi_{1,2}=\hbox{Linear Span}\{\sigma_{\alpha\beta}\vert\psi\rangle\,:\, \sigma_{\alpha\beta}\in K_{1,2}\}\ .
$$
Then, any vector orthogonal to $K_{1,2}^\psi$ is also orthogonal to $\vert\psi\rangle$.
\end{lemma}
\medskip

\proof
Consider the set $\{\sigma_{0i}\,,\,\sigma_{0j}\,,\,\sigma_{1k}\,,\,\sigma_{2k}\,,\,\sigma_{3k}\}$ and let $\vert0\rangle_2$ and $\vert1\rangle_2$ be the eigenvectors of $\sigma_k$. Furthermore, fix $i$ and $j$ so that
$$
\sigma_k\vert0\rangle_2=\vert0\rangle_2\ ,\ \sigma_k\vert1\rangle_2=\vert1\rangle_2\ ;\qquad \sigma_i\vert0\rangle_2=\vert1\rangle_2\ ,\ \sigma_i\vert1\rangle_2=\vert0\rangle_2\qquad\hbox{and}\qquad
\sigma_j\vert0\rangle_2=i\vert1\rangle_2\ ,\ \sigma_j\vert1\rangle_2=-i\vert0\rangle_2\ .
$$
Given $\vert\psi\rangle\,,\,\vert\varphi\rangle\in\CI^4$, consider their expansions with respect to the orthonormal basis $\{\vert ij\rangle=\vert i\rangle_1\otimes\vert j\rangle_2\}_{i,j=0}^1$ where
$\vert i\rangle_1$, $i=0,1$, are eigenvectors of $\sigma_3$:  $\vert\psi\rangle=\sum_{i,j=0}^1\psi_{ij}\vert ij\rangle$ and
$\vert\varphi\rangle=\sum_{i,j=0}^1\varphi_{ij}\vert ij\rangle$.
Acting with the first four $\sigma_{\alpha\beta}$ on $\vert\psi\rangle$, we have:
\begin{eqnarray}
\nonumber
\sigma_{0i}\vert\psi\rangle&=&\psi_{00}\vert 01\rangle+\psi_{01}\vert 00\rangle+\psi_{10}\vert 11\rangle+\psi_{11}\vert 10\rangle\ ,\quad
\sigma_{0j}\vert\psi\rangle=i(\psi_{00}\vert 01\rangle-\psi_{01}\vert 00\rangle+\psi_{10}\vert 11\rangle-\psi_{11}\vert 10\rangle)\\
\label{4vectors}
\sigma_{1k}\vert\psi\rangle&=&\psi_{00}\vert 10\rangle-\psi_{01}\vert 11\rangle+\psi_{10}\vert 00\rangle-\psi_{11}\vert 01\rangle\ ,\quad
\sigma_{2k}\vert\psi\rangle=i(\psi_{00}\vert 10\rangle-\psi_{01}\vert 11\rangle-\psi_{10}\vert 00\rangle+\psi_{11}\vert 01\rangle)\ .
\end{eqnarray}
If $\vert\varphi\rangle$ is orthogonal to each of the previous vectors, it follows that
\begin{eqnarray*}
\overline{\varphi_{01}}\psi_{00}+\overline{\varphi_{11}}\psi_{10}=0\ ,&&
\overline{\varphi_{00}}\psi_{01}+\overline{\varphi_{10}}\psi_{11}=0\\
\overline{\varphi_{10}}\psi_{00}-\overline{\varphi_{11}}\psi_{01}=0\ ,&&
\overline{\varphi_{00}}\psi_{10}-\overline{\varphi_{01}}\psi_{11}=0\ .
\end{eqnarray*}
These relations recast in matrix equation read
$$
\begin{bmatrix}
  0 & \psi_{00} & 0 & \psi_{10}\\
  \psi_{01} & 0 & \psi_{11} & 0\\
  0 & 0 & \psi_{00} & -\psi_{01}\\
  \psi_{10} & -\psi_{11} & 0 & 0
 \end{bmatrix}
 \begin{bmatrix}
  \varphi_{00} \\
  \varphi_{01}  \\
  \varphi_{10} \\
   \varphi_{11}
 \end{bmatrix}
=0\ .
$$
The determinant of the matrix is $2\psi_{00}\psi_{01}\psi_{10}\psi_{11}$; thus, $\vert\varphi\rangle$ orthogonal to the linear span of the vectors in~(\ref{4vectors}) can only exist if all components of $\vert\psi\rangle$ are non-zero. Then,
$$
\overline{\varphi_{01}}=-\frac{\overline{\varphi_{11}}\psi_{10}}{\psi_{00}}\ ,\ \overline{\varphi_{00}}=\frac{\overline{\varphi_{11}}\psi_{11}}{\psi_{00}}\ ,\ \overline{\varphi_{10}}=\frac{\overline{\varphi_{11}}\psi_{01}}{\psi_{00}}\quad\hbox{and}\quad
\vert\varphi\rangle=\overline{\psi_{11}}\vert 00\rangle+\overline{\psi_{10}}\vert 01\rangle-\overline{\psi_{01}}\vert 10\rangle-\overline{\psi_{00}}\vert 1\rangle\ .
$$
Such a vector results orthogonal to both the fifth vector $\sigma_{3k}\vert\psi\rangle$ and $\vert\psi\rangle$ itself.
Similar considerations hold for the set $\{\sigma_{i0}\,,\,\sigma_{j0}\,,\,\sigma_{k1}\,,\,\sigma_{k2}\,,\,\sigma_{k3}\}$.
\qed
\medskip

The next and final lemma concerns the dimensionality of the linear spans of vectors of the form $\sigma_{\alpha\beta}\vert\psi\rangle$ where the matrices belong to the set $K$ in Corollary~\ref{lastcor}, where three $\sigma$'s anti-commute and the fourth one commutes with only one of them.
\medskip

\begin{lemma}
\label{lem3}
Consider a sub-set $K=\{\sigma_{\alpha_{i}\beta_{i}}:i=1,...,4\}$ consisting of three anti-commuting matrices $\sigma_{\alpha\beta}$ plus a fourth one which commutes with only one of these three, say the first:
\begin{equation}
\label{Lindalg}
\{\sigma_{\alpha_{i}\beta_{i}}\,,\,\sigma_{\alpha_{j}\beta_{j}}\}=0\ , \ i,j=1,3\quad
[\sigma_{\alpha_{4}\beta_{4}}\,,\,\sigma_{\alpha_1\beta_1}]=0\ .
\end{equation}
Then, unless $\vert\psi\rangle\in\CI^4$ is an eigenstate of some $\sigma_{\mu\nu}$, $V_K=\CI^4$, where $V_K$ is the linear span of $\{\sigma_{\alpha_i\beta_i}\vert\psi\rangle\}_{i=1}^4$.
\end{lemma}
\medskip

\proof
Suppose $\vert\psi\rangle$ is not an eigenstate of any $\sigma_{\mu\nu}$, then the vectors
$\sigma_{\alpha_{1}\beta_{1}}\vert\psi\rangle$, $\sigma_{\alpha_{2}\beta_{2}}\vert\psi\rangle$ cannot be proportional.
For the same reason, the vectors $\sigma_{\alpha_{1}\beta_{1}}\vert\psi\rangle$, $\sigma_{\alpha_{2}\beta_{2}}\vert\psi\rangle$ and $\sigma_{\alpha_{4}\beta_{4}}\vert\psi\rangle$ are linearly independent. Indeed, suppose
$$
\sigma_{\alpha_{4}\beta_{4}}\vert\psi\rangle=\alpha\sigma_{\alpha_{1}\beta_{1}}\vert\psi\rangle +  \beta\sigma_{\alpha_{1}\beta_{1}}\vert\psi\rangle\ .
$$
Then, acting on both sides with $\sigma_{\alpha_{4}\beta_{4}}$ we obtain
$\vert\psi\rangle=\alpha\,S_1\vert\psi\rangle +  \beta\,S_2\vert\psi\rangle$, where
$S_i=\sigma_{\alpha_4\beta_4}\sigma_{\alpha_i\beta_i}$, $i=1,2$.
When substituting this expression for $\vert\psi\rangle$ in the right hand side of the equality,
the relations~(\ref{Lindalg}) yield
$$
\vert\psi\rangle=(\alpha^{2}-\beta^{2})\vert\psi\rangle+ \alpha\beta\,[\sigma_{\alpha_1\beta_1}\,,\,\sigma_{\alpha_2\beta_2}]\vert\psi\rangle\ ,
$$
which, for non-trivial $\alpha\beta$, is only possible if $\vert\psi\rangle$ is an eigenstate of the
the $\sigma_{\mu\nu}$ proportional to the commutator in the previous equality.
The same conclusions can be drawn about the linear independence of the four vectors spanning $V_K$.
Indeed, if
$$
\sigma_{\alpha_{3}\beta_{3}}\vert\psi\rangle= \alpha\sigma_{\alpha_{1}\beta_{1}}\vert\psi\rangle + \beta \sigma_{\alpha_{2}\beta_{2}}\vert\psi\rangle+ \gamma \sigma_{\alpha_{4}\beta_{4}}\vert\psi\rangle\ ,
$$
the same argument of before obtains
$$
\vert\psi\rangle=-(\alpha^2+\beta^2+\gamma^2)\vert\psi\rangle\,+\,\alpha\gamma\,
\{\sigma_{\alpha_1\beta_1}\,,\,\sigma_{\alpha_4\beta_4}\}\vert\psi\rangle\ .
$$
This implies that, unless $\vert\psi\rangle$ is eigenstate of the $\sigma_{\mu\nu}$ arising from the anti-commutator in the above expression, the four vectors spanning $V_K$ are linearly independent.
\qed
\medskip

The previous results show that, as a consequence of the assumption that no special quadruple $Q\in\cQ_{00}$ is contained in $I$, given a vector $\vert\psi\rangle\in\CI^4$ the sub-space linearly generated by $\sigma_{\alpha\beta}\vert\psi\rangle$ with $(\alpha\beta)\in I^c$ is either $\CI^4$ or contains $\vert\psi\rangle$.
The proof of Theorem~\ref{theo2} is thus completed by using Lemma~\ref{lem1} and its corollary.

Unfortunately, no further PPT entangled states different from those detected by the methods of Propositions~\ref{PPT2} and~\ref{PPT3} belong to the class described in Theorem~\ref{theo2}.

\begin{proposition}
PPT entangled lattice states satisfying the hypotheses of Propositions~\ref{PPT2} and~\ref{PPT3} also satisfy the hypothesis of Theorem~\ref{theo2}; vice versa, PPT lattice states that satisfy the hypothesis of Theorem~\ref{theo2} also satisfy those of Propositions~\ref{PPT2} and~\ref{PPT3}.
\end{proposition}

\proof
We start with the following observation: according to Lemma~\ref{lemquadr1} and Remark~\ref{remquadr}, one can exchange rows and columns mapping special quadruples into special quadruples  without altering the entanglement properties of the lattice states; it also follows that, by sending a lattice state $\rho_I$ into $\rho_{\tilde{I}}$ via such local unitary actions, the new pattern, $\tilde{I}$, has the same properties of the old one, $I$, in relation to whether the hypotheses of Propositions~\ref{PPT2} and~\ref{PPT3} and of Theorem~\ref{theo2} are fulfilled or not.
For instance, if there is a point $(\alpha,\beta)\in I$ none of whose special quadruples is contained in $I$, by sending $(\alpha,\beta)$ into $(\tilde{\alpha},\tilde{\beta})$ via a local unitary action, the same will be true of $(\tilde{\alpha},\tilde{\beta})\in\tilde{I}$ with respect to $\tilde{I}$: none of its special quadruples will be contained in $\tilde{I}$.
Also, if there is a row $R_\beta$ and a column $C_\alpha$ which contribute with only one point to $I$,  a part possibly from $(\alpha,\beta)$, then by sending $(\alpha,\beta)$ into $(\tilde{\alpha},\tilde{\beta})$ via a local unitary action, the same will be true of the transformed row and column $R_{\tilde{\beta}}$ and $C_{\tilde{\alpha}}$ with respect to $\tilde{I}$: there will be only one point on $R_{\tilde{\beta}}$ and $C_{\tilde{\alpha}}$, a part possibly from $(\tilde{\alpha},\tilde{\beta})$, contributing to $\tilde{I}$.

Now, suppose a PPT lattice state $\rho_I$ satisfies the requests of either Proposition~\ref{PPT2} or~\ref{PPT3}; then, there exists a column $C_\alpha$ and a row $R_\beta$ such that they contribute to $I$ with one point only, apart, possibly, for the point $(\alpha,\beta)$.
Then, by a suitable local rotation, this point can always be sent to $(0,0)\in I$ and the column $C_\alpha$ into the the column $C_0$, while the row $R_\beta$ becomes a row $R_j$ , $j\neq0$. The only contribution to $I$ from $C_0$ and $R_j$, apart, possibly, from the intersection point $(0,j)$, is $(0,0)$; thus, all other five points on these column and row do not contribute to $I$. The set of these five points correspond to anti-commuting $\sigma_{\alpha\beta}$; indeed, they are  of the form
$\sigma_{0i}$ and $\sigma_{ij}$ with $i\neq j$.
From the properties of the special quadruples in Lemma~\ref{lem0}, it follows that the absence from $I$ of this set of $5$ points forbids all special quadruples $Q_{00}$ to be contained in $I$. Therefore, according to Theorem~\ref{theo2}, the corresponding lattice state is entangled.

The second part of the proposition is proved by contradiction; namely, we show that if a lattice state $\rho_I$, with at least one point $(\alpha_0,\beta_0)\in I$ none of whose special quadruples is contained in $I$, violates the hypotheses of either Proposition~\ref{PPT2} or~\ref{PPT3}, then it must be NPT.
Because of the observation at the begnning of the proof, we can modify the pattern $I$ by local unitaries and send $(\alpha_0\,\beta_0)$ into $(0,0)$ while preserving the given constraints. We shall denote again by $I$ the transformed pattern.
Violation of the hypotheses of either Proposition~\ref{PPT2} or~\ref{PPT3} thus means that each pair consisting of the row $R_0$ and the column $C_j$, $j\neq 0$, respectively  of the column $C_0$ and the row $R_j$, $j\neq 0$, must contribute to $I$ with at least two points different from $(0,j)$, respectively $(j,0)$.

Let us consider the special quadruples in~(\ref{specQ00}); they can be divided into two groups: a group of nine special quadruples with at least one point on $C_0$ or $R_0$ and the group of six special quadruples with no point on this row and column (apart from $(0,0)$). These are explictily shown in~(\ref{mincov0}) and~(\ref{mincov1}) : none of them can be contained in $I$.
Since each point (apart from the origin) appears in exactly two of them, we could avoid all of them from being contained in $I$, by excluding three points to contribute to $I$. However, this would result in $I$ with an empty column or an empty row thus contradicting the fact that $\rho_I$ violates the hypotheses of Propositions~\ref{PPT2} and~\ref{PPT3}.
Therefore, at leat one $R_i$ and $C_j$ must contribute to $I$ with at least two points: patterns fulfilling the hypothesis of Theorem~\ref{theo2} and violate those of Propositions~\ref{PPT2} and~\ref{PPT3} can be derived from
$$
 \begin{array}{c|c|c|c|c}
               3 & \quad\!   & \times   &  \circ &  \circ    \\
               \hline
               2 & \quad\!  & \times & \circ  &  \circ  \\
               \hline
               1 &  \quad\!  & \quad\! & \times & \times\\
               \hline
               0 & \times & &  &      \\
               \hline
                 & 0 & 1 & 2 & 3
 \end{array} \quad,
$$
where the circles denote sites that cannot contribute to $I$, while the empty squares may still contribute.
Indeed, by making use of local rotations that leave $(0,0)$ invariant, one shows that all patterns that fulfill the constraints must contain sub-patterns that follow from the two previous one by suitable exchanges of rows and columns.

As a consequence of Proposition~\ref{PPT}, if $\rho_I$ is to be PPT, the number of contributing sites must be $N_I\geq 8$; this means that we have to accommodate at least three more crosses still avoiding special quadruples in $\cQ_{00}$ from being contained in $I$. All patterns achieving this with the least number of avoded sites come, by local rotations, from
$$
 \begin{array}{c|c|c|c|c}
               3 &    & \times   &  \circ &  \circ    \\
               \hline
               2 &   & \times & \circ  &  \circ  \\
               \hline
               1 &  \circ  & \quad\! & \times & \times\\
               \hline
               0 & \times &\circ &  &\quad\!      \\
               \hline
                 & 0 & 1 & 2 & 3
 \end{array} \quad .
$$
Accomodating other three points implies that at least one row and one column contribute with $5$ points; therefore, in order to guarantee PPT-ness, $N_I$ must be at least $10$; this fixes the source pattern to be
\begin{equation*}
 \begin{array}{c|c|c|c|c}
               3 &\times    & \times   &  \circ &  \circ    \\
               \hline
               2 &   \times& \times & \circ  &  \circ  \\
               \hline
               1 &  \circ  & \times & \times & \times\\
               \hline
               0 & \times &\times &\times  &\circ     \\
               \hline
                 & 0 & 1 & 2 & 3
 \end{array} \quad .
 \end{equation*}
that  corresponds to an NPT state for there are rows and columns contributing with $6>N_I/2=5$ points.
\qed

\section{Conclusions}

In this work we have considered a particular class of bipartite states, called lattice states, of two pairs of two qubits described by
density matrices $\rho_I$  acting on $\CI^4\otimes\CI^4$~\cite{benatti0}.
They are uniform mixtures of orthogonal projections indexed by points $(\alpha,\beta)$  belonging to subsets $I$ of the finite square lattice of cardinality $16$. The projections are generated by the action of $\1_4\otimes\sigma_{\alpha\beta}$ on the completely symmetric vector in $\CI^4\otimes\CI^4$, where $\sigma_{\alpha\beta}$ is the tensor product of the Pauli matrices $\sigma_{\alpha}$, $\sigma_\beta$ .
We have generalized them to states on $\CI^{2^n}\otimes\CI^{2^n}$, referred to as $\sigma-$diagonal states.

One of the main issues was to decide whether a given $\sigma-$diagonal state, is entangled or separable. We have tackled this problem using the results of~\cite{Stoermer} and showed that, starting from a general non-diagonal positive map, possible entanglement of $\sigma-$diagonal states would be revealed using only a particularly simple sub-class of positive maps adapted to the diagonal structure of the states, that are combinations of elementary maps of the form $X\mapsto \sigma_{\alpha\beta}\,X\,\sigma_{\alpha\beta}$.

Concerning the lattice states in $\CI^4\otimes\CI^4$, we have included the results given in~\cite{{benatti1},{benatti2}} in this more general framework and investigated the relations between  entanglement $($respectively, separability$)$ of the lattice states and  the geometrical patterns that identify them.
The techniques we have developed to this task are based on particular separable lattice states consisting of four points, called special quadruples, which play a crucial role in this game.
Using the notion of uniform covering by special quadruples contained in the subset $I$, we could show that some of the lattice states whose separability was not known from previous works, are indeed separable.
Also, by constructing a proper entanglement witness, we have shown that if there exists a point in the set $I$ such that none of its special quadruples are contained in $I$, then the corresponding lattice state $\rho_I$, is entangled.

In view of these results, we formulate the following conjecture: entangled lattice states $\rho_I$ are either NPT or they posses at least one point $(\alpha,\beta)\in I$ with no special quadruple $Q_{\alpha\beta}$ contained in $I$.
In order to prove this conjecture, we have to control the properties of PPT lattice states whose separability could not be decided by the methods developed in this work; for instance,
$$
 \begin{array}{c|c|c|c|c}
               3 & \times  & \quad\!   &  \quad\! &  \times   \\
               \hline
               2 & \times  & \times & \quad\!  &  \times   \\
               \hline
               1 &  \times  & \times   & \times &  \quad\!    \\
               \hline
               0 & \times & \times & \times &      \\
               \hline
                 & 0 & 1 & 2 & 3
 \end{array}\quad .
 $$
All points of this PPT lattice state of rank $11$ have at least three special quadruples contained in I, but no uniform covering could be found so far.
This problem will be tackled in a subsequent work.

\section{Appendix A}

The patterns corresponding to the special quadruples (the point $(0,0)$ common to all of them has been omitted)
\begin{eqnarray*}
\nonumber
&&\hskip-.5cm
\{(0,1);(1,0);(1,1)\}  \quad     \{(0,2);(2,0);(2,2)\} \quad   \{(0,3);(3,0);(3,3)\}
\quad \{(1,1);(2,2);(3,3)\} \quad \{(1,2);(2,3);(3,1)\}\\
\nonumber
&&\hskip-.5cm
\{(0,1);(2,1);(2,0)\} \quad \{(0,2);(1,2);(1,0)\}  \quad \{(0,3);(1,3);(1,0)\}
\quad \{(1,1);(2,3);(3,2)\} \quad \{(1,3);(2,2);(3,1)\}\\
&&\hskip-.5cm
\{(0,1);(3,1);(3,0)\}   \quad    \{(0,2);(3,2);(3,0)\}  \quad  \{(0,3);(2,3);(2,0)\}
\quad \{(1,2);(2,1);(3,3)\}  \quad \{(1,3);(2,1);(3,2)\}
\end{eqnarray*}
divide into two families: the first three columns correspond to rectangular patterns:
\begin{equation}
\label{mincovrect1}
\begin{array}{c|c|c|c|c}
               3 &  \quad\!  &  \quad\!  &  \quad\! &  \quad\!    \\
               \hline
               2 & \quad\!  & \quad\! & \quad\!  &  \quad\!    \\
               \hline
               1 &  \times  &  \times  &  \quad\!  &  \quad\!     \\
               \hline
               0 & \times & \times  & \quad\! &   \quad\!   \\
               \hline
                 & 0 & 1 & 2 & 3
 \end{array}\quad ,\quad
\begin{array}{c|c|c|c|c}
               3 & \quad\!   & \quad\!   &  \quad\! &  \quad\!    \\
               \hline
               2 & \quad\! & \quad\! & \quad\!  &  \quad\!    \\
               \hline
               1 &  \times  & \quad\!   & \times &  \quad\!     \\
               \hline
               0 & \times & \quad\!  & \times &\quad\!      \\
               \hline
                 & 0 & 1 & 2 & 3
 \end{array}\quad ,\quad
\begin{array}{c|c|c|c|c}
               3 & \quad\!   & \quad\!   &  \quad\! &  \quad\!    \\
               \hline
               2 & \quad\!  & \quad\! & \quad\!  &  \quad\!    \\
               \hline
               1 &  \times  & \quad\!   & \quad\! &  \times     \\
               \hline
               0 & \times & \quad\!  & \quad\! &\times      \\
               \hline
                 & 0 & 1 & 2 & 3
 \end{array}
\end{equation}
\begin{equation}
\label{mincovrect2}
\begin{array}{c|c|c|c|c}
               3 & \quad\!   & \quad\!   &  \quad\! &  \quad\!    \\
               \hline
               2 & \times & \quad\! & \times &  \quad\!    \\
               \hline
               1 &  \quad\!  & \quad\!   & \quad\! &  \quad\!     \\
               \hline
               0 & \times & \quad\!  & \times &\quad\!      \\
               \hline
                 & 0 & 1 & 2 & 3
\end{array}\quad ,\quad
\begin{array}{c|c|c|c|c}
               3 &  \quad\!  &  \quad\!  &  \quad\! &  \quad\!    \\
               \hline
               2 & \times  & \times & \quad\!  &  \quad\!    \\
               \hline
               1 &  \quad\!  &  \quad\!  &  \quad\!  &  \quad\!     \\
               \hline
               0 &\times & \times  & \quad\! &      \\
               \hline
                 & 0 & 1 & 2 & 3
 \end{array}\quad ,\quad
\begin{array}{c|c|c|c|c}
               3 & \quad\!   & \quad\!   &  \quad\! &  \quad\!    \\
               \hline
               2 & \times  & \quad\! & \quad\!  &  \times    \\
               \hline
               1 &  \quad\!  & \quad\!   & \quad\! &  \quad\!     \\
               \hline
               0 & \times & \quad\!  & \quad\! &\times      \\
               \hline
                 & 0 & 1 & 2 & 3
 \end{array}
\end{equation}
\begin{equation}
\label{mincovrect3}
\begin{array}{c|c|c|c|c}
               3 & \times   & \quad\!   &  \quad\! &  \times    \\
               \hline
               2 & \quad\!  & \quad\! & \quad\!  &  \quad\!    \\
               \hline
               1 &  \quad\!  & \quad\!   & \quad\! &  \quad\!     \\
               \hline
               0 & \times & \quad\!  & \quad\! &\times      \\
               \hline
                 & 0 & 1 & 2 & 3
 \end{array}\quad ,\quad
\begin{array}{c|c|c|c|c}
               3 &  \times  &  \times  &  \quad\! &  \quad\!    \\
               \hline
               2 & \quad\!  & \quad\! & \quad\!  &  \quad\!    \\
               \hline
               1 &  \quad\!  &  \quad\!  &  \quad\!  &  \quad\!     \\
               \hline
               0 & \times & \times  & \quad\! & \quad\!     \\
               \hline
                 & 0 & 1 & 2 & 3
 \end{array}\quad ,\quad
\begin{array}{c|c|c|c|c}
               3 & \times  & \quad\!   &  \times &  \quad\!    \\
               \hline
               2 & \quad\! & \quad\! & \quad\!  &  \quad\!    \\
               \hline
               1 &  \quad\!  & \quad\!   & \quad\! &  \quad\!     \\
               \hline
               0 & \times & \quad\!  & \times &\quad\!      \\
               \hline
                 & 0 & 1 & 2 & 3
 \end{array}
\quad .
\end{equation}
The second family corresponding to the last $6$ quadruples presents the following patterns:
\begin{equation}
\begin{array}{c|c|c|c|c}
               3 & \quad\! & \quad\! & \quad\! &\times   \\
               \hline
               2 & \quad\! & \quad\! &  \times &\quad\!  \\
               \hline
               1 & \quad\! & \times& \quad\!  &\quad\!  \\
               \hline
               0 & \times & \quad\! & \quad\!  &   \\
               \hline
                 & 0 & 1 & 2 & 3
\end{array}\quad ,\quad
\begin{array}{c|c|c|c|c}
               3 & \quad\! & \quad\! & \times &\quad\!   \\
               \hline
               2 & \quad\! & \quad\! &\quad\! &\times  \\
               \hline
               1 & \quad & \times & \quad\!  &\quad\!   \\
               \hline
               0 & \times & \quad\! & \quad\!  &   \\
               \hline
                 & 0 & 1 & 2 & 3
\end{array}\quad\ ,\quad
\begin{array}{c|c|c|c|c}
               3 & \quad\! & \quad\! & \quad\! &\times   \\
               \hline
               2 & \quad\! & \times &  \quad\! & \quad\! \\
               \hline
               1 & \quad\! & \quad\! & \times  & \quad\!  \\
               \hline
               0 & \times & \quad\! & \quad\!  &   \\
               \hline
                 & 0 & 1 & 2 & 3
 \end{array}
\label{mincov0}
\end{equation}
\begin{equation}
\label{mincov1}
\begin{array}{c|c|c|c|c}
               3 & \quad\! & \quad\! & \times &\quad\!   \\
               \hline
               2 & \quad\! & \times &  \quad\! & \quad\! \\
               \hline
               1 & \quad\! & \quad\! & \quad\!  & \times  \\
               \hline
               0 & \times & \quad\! & \quad\!  &   \\
               \hline
                 & 0 & 1 & 2 & 3
 \end{array}
\quad ,\quad
 \begin{array}{c|c|c|c|c}
               3 & \quad\! & \times & \quad\! &\quad\!   \\
               \hline
               2 & \quad\! & \quad\! &  \times & \quad\! \\
               \hline
               1 & \quad\! & \quad\! & \quad\!  & \times  \\
               \hline
               0 & \times & \quad\! & \quad\!  &   \\
               \hline
                 & 0 & 1 & 2 & 3
 \end{array}
\quad ,\quad
\begin{array}{c|c|c|c|c}
               3 & \quad\! & \times & \quad\!&\quad\!   \\
               \hline
               2 & \quad\! & \quad\! &  \quad\! & \times \\
               \hline
               1 & \quad & \quad\! & \times  &\quad\!   \\
               \hline
               0 & \times & \quad\! & \quad\!  &   \\
               \hline
                 & 0 & 1 & 2 & 3
\end{array}\quad .
\end{equation}
The two families are singularly mapped into themselves by local unitary operations as in Lemma~\ref{lemquadr1} and Remark~\ref{remquadr}.

As stated in Lemma \ref{lem0} the special quadruples in $\cQ_{00}$ enjoy the following properties:
\begin{enumerate}
\item
two different special quadruples cannot have more than one lattice point in common;
\item
lattice points belonging to different quadruples with a point (different from $(0,0)$) in common correspond to anti-commuting $\sigma$'s;
\item
each lattice point belongs to three different special quadruples;
\item
let $(\alpha_1,\beta_1)$ and $(\alpha_2,\beta_2)$ be such that $\sigma_{\alpha_1\beta_1}$ and $\sigma_{\alpha_2\beta_2}$
anti-commute and consider the three special quadruples with  $(\alpha_1,\beta_1)$, respectively  $(\alpha_2,\beta_2)$  in common.
They consist of $9$ lattice lattice points different from $(0,0)$,  $(\alpha_1,\beta_1)$ and $(\alpha_2,\beta_2)$ , which divide into three disjoint subsets corresponding to anti-commuting $\sigma$'s.
\end{enumerate}

\noindent
\proof
The first two properties follow from the fact that two commuting $\sigma_{\alpha\beta}\neq\sigma_{00}$ and  $\sigma_{\gamma\delta}\neq\sigma_{00}$  determine a unique $\sigma_{\mu\nu}\neq\sigma_{00}$ commuting  with both of them.\\
The third property descends from the first and the fact that each $\sigma_{\alpha\beta}$ commutes with other $6$ $\sigma$'s.\\
The last property is a consequence of the fact that the $3$ special quadruples with the point $(\alpha_1,\beta_1)$ in common and the three special quadruples with $(\alpha_2,\beta_2)$  in common can be grouped in pairs of special quadruples that share one point. For instance, consider $(1,2)$ and $(1,3)$;
the special quadruples relative to them are
\begin{eqnarray*}
\{(0,0),(1,2),(1,0),(0,2)\}\quad, &&\quad\{(0,0),(1,3),(1,0),(0,3)\}\\
\{(0,0),(1,2),(2,1),(3,3)\}\quad, &&\quad\{(0,0),(1,3),(2,2),(3,1)\}\\
\{(0,0),(1,2),(2,3),(3,1)\}\quad, &&\quad\{(0,0),(1,3),(2,1),(3,2)\}\qquad.
\end{eqnarray*}
The first one in the left column has the point $(1,0)$ in common with the first one in the right column; the second one in the left column has the point $(2,1)$ in common with the third one in the right column; finally, the third one in the left column has the point $(3,1)$ in common with the second one in the right column. The $3$ sets corresponding to anti-commuting $\sigma$'s are then given by the following columns
\begin{eqnarray*}
&&(1,0)\quad,\quad(0,2)\quad,\quad(0,3)\\
&&(2,1)\quad,\quad(3,3)\quad,\quad(2,2)\\
&&(3,1)\quad,\quad(2,3)\quad,\quad(3,2)\qquad.
\end{eqnarray*}
\qed

\section{Appendix B}

Besides the state in~(\ref{Ex10}), also those corresponding to the following  following patterns can be shown to be separable by means of  Proposition~\ref{theo3}:
$$
\underbrace{ \begin{array}{c|c|c|c|c}
               3 & \quad\!  & \times  & \times   & \times   \\
               \hline
               2 & \quad\!  & \times  & \quad\!   & \times \\
               \hline
               1 &  \quad\!  & \times  & \times   & \times   \\
               \hline
               0 & \times & \quad\! & \quad\!     \\
               \hline
                 & 0 & 1 & 2 & 3
 \end{array}}_{I_1}\quad,\quad
  \underbrace{\begin{array}{c|c|c|c|c}
               3 & \quad\!  & \times  & \times   & \times   \\
               \hline
               2 & \quad\!  & \times  & \quad\!  & \times \\
               \hline
               1 &  \quad\!  & \times  & \quad\!  & \times   \\
               \hline
               0 & \times & \quad\! & \quad\!     \\
               \hline
                 & 0 & 1 & 2 & 3
 \end{array}}_{I_2}\,.
$$
\medskip
Indeed, the lattice states $\rho_{I_{1,2}}$ can be convexly decomposed as follows (we identify them with the special quadruples of the uniform coverings of $I_{1,2}$):
\begin{eqnarray*}
  \begin{array}{c|c|c|c|c}
               3 & \quad\!  & \times & \times  & \times     \\
               \hline
               2 & \quad\!  &  \times   &  \quad\!  &  \times   \\
               \hline
               1 &  \quad\! & \times &  \quad\! & \times     \\
               \hline
               0 & \times   &     &     &      \\
               \hline
                 & 0 & 1 & 2 & 3
 \end{array}\quad&=&\frac{1}{2}\quad
 \begin{array}{c|c|c|c|c}
               3 & \quad\!   & \quad\!   &  \times &  \quad\!    \\
               \hline
               2 & \quad\!  & \quad\! & \quad\!  &  \times    \\
               \hline
               1 &  \quad\!  & \times   & \quad\! &  \quad\!     \\
               \hline
               0 & \times & \quad\!  & \quad\! &      \\
               \hline
                 & 0 & 1 & 2 & 3
 \end{array}\quad +\frac{1}{2}\quad
 \begin{array}{c|c|c|c|c}
               3 & \quad\!   & \quad\!   &  \times &  \quad\!    \\
               \hline
               2 & \quad\!  & \times & \quad\!  &  \quad\!    \\
               \hline
               1 &  \quad\!  & \quad\!   & \quad\! &  \times    \\
               \hline
               0 & \times & \quad\!  & \quad\! &      \\
               \hline
                 & 0 & 1 & 2 & 3
 \end{array}
 +\frac{1}{2}\quad
 \begin{array}{c|c|c|c|c}
               3 & \quad\!   & \times   &  \quad\! &  \times    \\
               \hline
               2 & \quad\!  & \quad\! & \quad\!  &  \quad\!    \\
               \hline
               1 &  \quad\!  & \times   & \quad\! &  \times    \\
               \hline
               0 & \quad\! & \quad\!  & \quad\! &      \\
               \hline
                 & 0 & 1 & 2 & 3
 \end{array}\quad +\frac{1}{2}\quad
 \begin{array}{c|c|c|c|c}
               3 & \quad\!   & \times   &  \quad\! &  \times    \\
               \hline
               2 & \quad\!  & \times & \quad\!  &  \times   \\
               \hline
               1 &  \quad\!  & \quad\!   & \quad\! &  \quad\!     \\
               \hline
               0 & \quad\! & \quad\!  & \quad\! &      \\
               \hline
                 & 0 & 1 & 2 & 3
 \end{array}\\
\\
\begin{array}{c|c|c|c|c}
               3 & \quad\!   & \times   &  \times &  \times    \\
               \hline
               2 & \quad\!  & \times & \quad\!  &  \times    \\
               \hline
               1 &  \quad\!  & \times   & \times &  \times     \\
               \hline
               0 & \times & \quad\!  & \quad\! &      \\
               \hline
                 & 0 & 1 & 2 & 3
 \end{array}\quad &=&\frac{1}{4}\quad
 \begin{array}{c|c|c|c|c}
               3 & \quad\!   & \quad\!   &  \times &  \quad\!    \\
               \hline
               2 & \quad\!  & \times & \quad\!  &  \times   \\
               \hline
               1 &  \quad\!  & \times & \quad\! & \quad\!\\
               \hline
               0 & \times & \quad\!  & \quad\! &  \quad\!    \\
               \hline
                 & 0 & 1 & 2 & 3
 \end{array} \quad +\frac{1}{4}\quad
 \begin{array}{c|c|c|c|c}
               3 & \quad\!   & \times   &  \quad\! &  \quad\!    \\
               \hline
               2 & \quad\!  & \quad\! & \quad\!  &  \times   \\
               \hline
               1 &  \quad\!  & \quad\!   & \times &  \quad\!     \\
               \hline
               0 & \times & \quad\!  & \quad\! &      \\
               \hline
                 & 0 & 1 & 2 & 3
 \end{array}
 +\frac{1}{4}\quad
 \begin{array}{c|c|c|c|c}
               3 & \quad\!   & \quad\!   &  \quad\! &  \times    \\
               \hline
               2 & \quad\!  & \times & \quad\!  &  \quad\!   \\
               \hline
               1 &  \quad\!  & \quad\!   & \times &  \quad\!     \\
               \hline
               0 & \times & \quad\!  & \quad\! &      \\
               \hline
                 & 0 & 1 & 2 & 3
 \end{array}\quad +\frac{1}{4}\quad
 \begin{array}{c|c|c|c|c}
               3 & \quad\!   & \quad\!   &  \times &  \quad\!    \\
               \hline
               2 & \quad\!  & \times & \quad\!  &  \quad\!   \\
               \hline
               1 &  \quad\!  & \quad\!   & \quad\! &  \times     \\
               \hline
               0 & \times & \quad\!  & \quad\! &      \\
               \hline
                 & 0 & 1 & 2 & 3
 \end{array}\\
 &+&\frac{1}{4}\quad
 \begin{array}{c|c|c|c|c}
               3 & \quad\!   & \times   &  \times &  \quad\!   \\
               \hline
               2 & \quad\!  & \quad\! & \quad\!  &  \quad\!   \\
               \hline
               1 &  \quad\!  & \times   & \times &  \quad\!     \\
               \hline
               0 & \quad\! & \quad\!  & \quad\! &      \\
               \hline
                 & 0 & 1 & 2 & 3
 \end{array}\quad +\frac{1}{4}\quad
 \begin{array}{c|c|c|c|c}
               3 & \quad\!   & \quad\!   &  \times &  \times    \\
               \hline
               2 & \quad\!  & \quad\! & \quad\!  &  \quad\!  \\
               \hline
               1 &  \quad\!  & \quad\!   & \times &  \times     \\
               \hline
               0 & \quad\! & \quad\!  & \quad\! &      \\
               \hline
                 & 0 & 1 & 2 & 3
 \end{array}
 +\frac{1}{4}\quad
 \begin{array}{c|c|c|c|c}
               3 & \quad\!   & \times   &  \quad\! &  \times    \\
               \hline
               2 & \quad\!  & \quad\! & \quad\!  &  \quad\!   \\
               \hline
               1 &  \quad\!  & \times   & \quad\! &  \times     \\
               \hline
               0 & \quad\! & \quad\!  & \quad\! &      \\
               \hline
                 & 0 & 1 & 2 & 3
 \end{array}\quad +\frac{1}{4}\quad
 \begin{array}{c|c|c|c|c}
               3 & \quad\!   & \quad\!   &  \quad\! &  \quad\!   \\
               \hline
               2 & \quad\!  & \times & \quad\!  &  \times   \\
               \hline
               1 &  \quad\!  & \times   & \quad\! &  \times    \\
               \hline
               0 & \quad\! & \quad\!  & \quad\! &      \\
               \hline
                 & 0 & 1 & 2 & 3
 \end{array}\\
 &+&
 \frac{1}{4}\quad\begin{array}{c|c|c|c|c}
               3 & \quad\!   & \times   &  \quad\! &  \times    \\
               \hline
               2 & \quad\!  & \times & \quad\!  &  \times   \\
               \hline
               1 &  \quad\!  & \quad\!   & \quad\! &  \quad\!     \\
               \hline
               0 & \quad\! & \quad\!  & \quad\! &      \\
               \hline
                 & 0 & 1 & 2 & 3
 \end{array}\ .
\end{eqnarray*}
All those lattice states that result from the previous ones by application of local unitary operations as in Lemma~\ref{lemquadr1} and Remark~\ref{remquadr} result separable as well.

\section*{Acknowledgement}
One  of us, MK, would like to express her gratitude to SISSA and in particular to Gianfausto dell'Antonio for his encouragement, support and scientific advice.

\end{document}